\documentclass[
 reprint,
 amsmath,amssymb,
 aps,
 prl,
 superscriptaddress
]{revtex4-2}

\usepackage[utf8]{inputenc}
\usepackage{graphicx}
\usepackage{hyperref}
\usepackage{siunitx}
\usepackage{gensymb}
\usepackage{textcomp}
\usepackage[version=4]{mhchem}
\newcommand{\figsize}{1.}

\begin{document}

\preprint{APS/123-QED}

\title{Optical Mass Spectrometry of Cold $\mathrm{RaOH}^+$ and ${\mathrm{RaOCH}_3}^+$}

\author{M. Fan}
\affiliation{Department of Physics, University of California, Santa Barbara, California 93106, USA}
\affiliation{California Institute for Quantum Entanglement, Santa Barbara, California 93106, USA}

\author{C. A. Holliman}
\affiliation{Department of Physics, University of California, Santa Barbara, California 93106, USA}
\affiliation{California Institute for Quantum Entanglement, Santa Barbara, California 93106, USA}

\author{X. Shi}
\affiliation{Department of Physics, University of California, Santa Barbara, California 93106, USA}
\affiliation{California Institute for Quantum Entanglement, Santa Barbara, California 93106, USA}

\author{H. Zhang}
\affiliation{CAS Key Laboratory of Quantum Information, University of Science and Technology of China, Hefei, 230026, China}

\author{M. W. Straus}
\affiliation{Department of Physics, University of California, Santa Barbara, California 93106, USA}
\affiliation{California Institute for Quantum Entanglement, Santa Barbara, California 93106, USA}

\author{X. Li}
\affiliation{Key Laboratory for Physical Electronics and Devices of the Ministry of Education and Shaanxi Key Laboratory of Information Photonic Technique, Xi\textquotesingle an Jiaotong University, Xi\textquotesingle an 710049, China}

\author{S. W. Buechele}
\affiliation{Department of Physics, University of California, Santa Barbara, California 93106, USA}
\affiliation{California Institute for Quantum Entanglement, Santa Barbara, California 93106, USA}

\author{A. M. Jayich}
\email{jayich@gmail.com}
\affiliation{Department of Physics, University of California, Santa Barbara, California 93106, USA}
\affiliation{California Institute for Quantum Entanglement, Santa Barbara, California 93106, USA}

\date{\today}

\begin{abstract}

We present an all-optical mass spectrometry technique to identify trapped ions. The new method uses laser-cooled ions to determine the mass of a cotrapped dark ion with a sub-dalton resolution within a few seconds. We apply the method to identify the first controlled synthesis of cold, trapped $\mathrm{RaOH}^+$ and ${\mathrm{RaOCH}_3}^+$.  These molecules are promising for their sensitivity to time and parity violations that could constrain sources of new physics beyond the standard model. The nondestructive nature of the mass spectrometry technique may help identify molecular ions or highly charged ions prior to optical spectroscopy.  Unlike previous mass spectrometry techniques for small ion crystals that rely on scanning, the method uses a Fourier transform that is inherently broadband and comparatively fast.    The technique's speed provides new opportunities for studying state-resolved chemical reactions in ion traps.

\end{abstract}

\maketitle

\textit{Introduction.}---Ion traps are powerful tools because their ability to trap only depends on two properties: the mass and charge of a particle.  Therefore they can trap ionic species with rich internal structures that preclude laser cooling or fluorescence.  Such dark ions include molecules, highly charged ions, and atoms with transitions that are deep in the UV.  These ions can be sympathetically cooled by cotrapped laser-cooled ions, where they appear as dark ion defects in a Coulomb crystal.  Dark ions have seen great successes in optical clocks,  e.g., \ce{Al+}\cite{Brewer2019}, constraining new physics,  e.g., \ce{HfF+}\cite{Cairncross2017} and studying state-resolved chemical reactions, e.g., \ce{BaCl+}\cite{Puri2019} and \ce{RbSr+}\cite{Sikorsky2018}.  There has also been much progress with highly charged ions, which generally lack strong fluorescence transitions, for metrology and tests of fundamental constant variations \cite{Safronova2014, Kozlov2018, Micke2020}. In this work we have synthesized a pair of molecular ions that are promising for probing new physics: \ce{RaOH+} and \ce{RaOCH3+}.

Recent measurements of parity ($P$) and time-reversal ($T$) violating moments are now probing physics at energy scales beyond the direct reach of the Large Hadron Collider \cite{Andreev2018}. Radium-based molecules are promising for constraining hadronic $P$, $T$-odd forces \cite{Flambaum2019, Kozyryev2017, GarciaRuiz2020}. The heavy and octupole-deformed radium nucleus enhances sensitivity to new physics in the hadronic sector \cite{Gaffney2013, Graner2016}.  This sensitivity is further enhanced when radium is incorporated into molecules such as \ce{RaOH+} or \ce{RaOCH3+} \cite{Kudashov2014, Kozyryev2017, Flambaum2019} that have large effective electric fields and molecular structure that is critical for reducing systematic uncertainties.   An ion trap is advantageous for working with radioactive molecules as high measurement sensitivity can be achieved with small sample sizes due to long measurement times \cite{Cairncross2017, Zhou2020}.  For example, the long trap times combined with the high sensitivity of \ce{^{225}RaOCH3+} are sufficient for an experiment with even a single trapped molecule to set new bounds on hadronic $P$, $T$ violations \cite{Yu2021}.

Because dark ions do not fluoresce, mass spectrometry techniques are commonly used for species identification.  We present a new nondestructive optical mass spectrometry (OMS) technique to identify a trapped dark ion in a Coulomb crystal by measuring a motional frequency of the crystal. In this work, we use cotrapped laser-cooled ions to amplify the secular (normal mode) motion of the crystal by using coherent population trapping (CPT) in the $S_{1/2}$ - $P_{1/2}$ - $D_{3/2}$ $\Lambda$-level system \cite{Chen2015} common to Ca$^+$, Sr$^+$, Ba$^+$, and Ra$^+$.  For these ions, it is fast and straightforward to realize CPT by changing the frequency and power of the $P_{1/2} \rightarrow D_{3/2}$ repump laser from the laser cooling values with an acousto-optical modulator.  With CPT, the optical spectrum of the $S_{1/2} \rightarrow P_{1/2}$ cooling transition can be modified so the ion's motion is coherently amplified \cite{Berkeland2002}, which modulates the scattered light at the motional frequency and its harmonics, which can then be measured with a Fourier transform. Because the motional modes of the ion crystal are set by the charge and mass of the trapped particles, motional frequencies can be used to determine the ion's mass. The OMS technique can be used with any laser-cooled ion, even without using a $\Lambda$ structure, e.g., Be$^+$, Mg$^+$, or Yb$^+$, via ``phonon lasing'' with bichromatic light \cite{Vahala2009}.

Many techniques have been used to identify trapped dark ions in Paul traps. The best technique for large Coulomb crystals (hundreds to thousands of ions) with multiple dark ion species is time-of-flight mass spectrometry, but it is inherently destructive and requires a purpose-built trap and custom electronics \cite{Schneider2014, Deb2015, Schmid2017}.  In the regime of small ion crystals with a few dark ions, multiple techniques have been developed that rely on measuring a trap secular frequency, including secular motion excitation by applied electric fields (tickle scans) \cite{Staanum2008, Willitsch2008, Gingell2010}, optical sideband spectroscopy \cite{Goeders2013a}, and ion crystal phase transitions \cite{Groot-Berning2019}.   Optical sideband spectroscopy requires a narrow linewidth laser and is also slow ($\sim1$ minute).  Measuring phase transitions has limited performance due to the large mass uncertainties
 ($\sim 5$ daltons \cite{Groot-Berning2019}). A tickle scan typically requires a $\geq 1$ minute measurement where an electrical drive is scanned over the secular frequency, which results in a broad resonance peak due to the damping from the laser cooling that is required by the technique. The applied electric field can also destructively drive ions out of the trap. Despite these drawbacks and the need for additional electrical connections (which are a noise conduit), for small crystals the tickle scan has been the most widely used technique. A variation of the tickle scan method modulates the cooling light intensity instead of modulating an electrical drive \cite{Drewsen2004}.  This removes some drawbacks but with an increase in technical overhead. In order for any of these techniques to achieve reasonable measurement times ($\sim$1 minute), \emph{a priori} knowledge of the trapped dark ions is required to reduce the secular frequency scan range.  In comparison to these small ion crystal mass spectrometry techniques, the reported OMS technique is faster, does not require knowledge of the dark ion's mass, is less invasive, and is a simple extension to Doppler cooling.

\begin{figure}
    \centering
    \includegraphics[width=\figsize\linewidth]{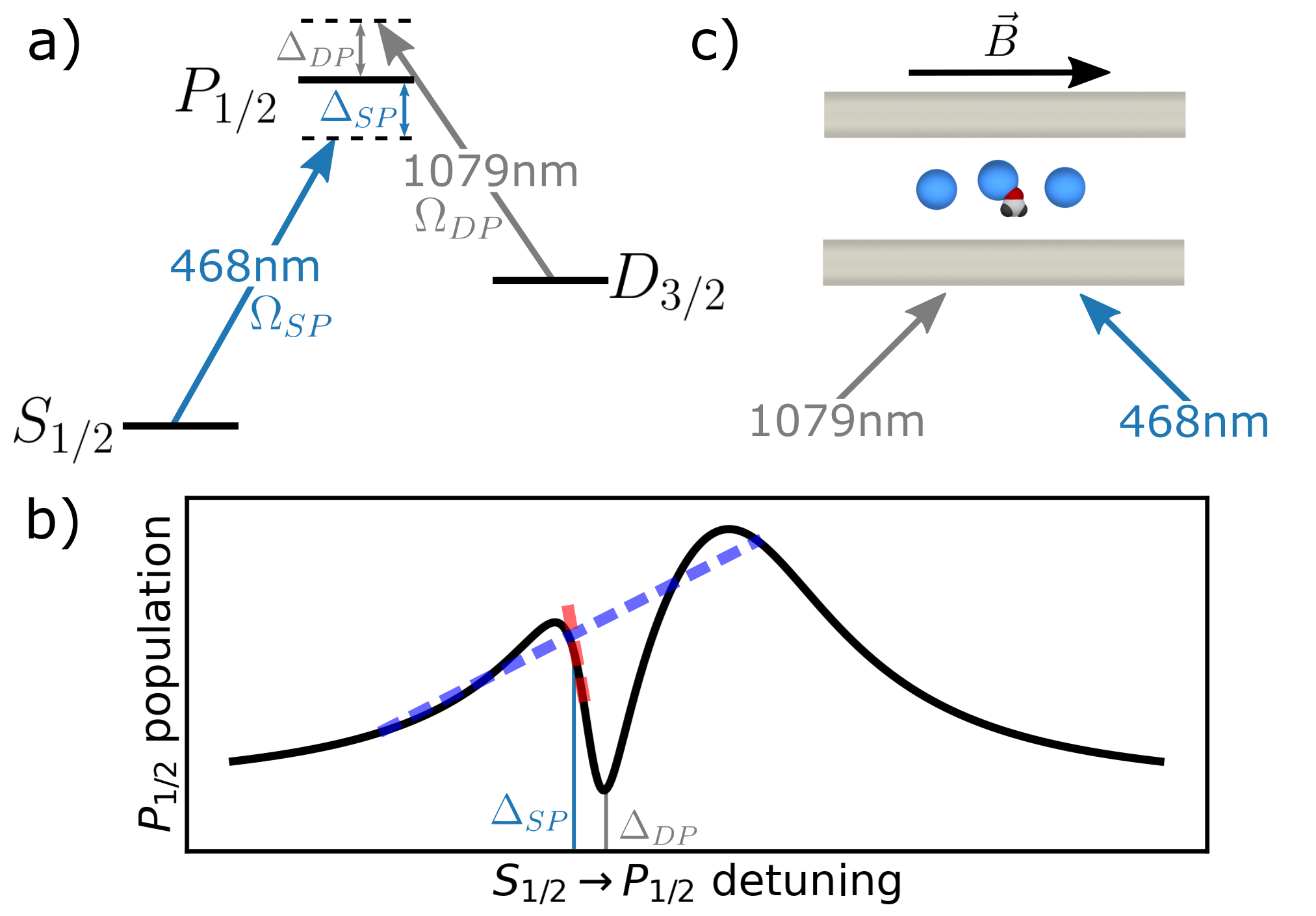}
    \caption{(a) \ce{Ra+} energy levels and transitions used in this work. $\Delta_{\mathrm{SP}}$ ($\Delta_{\mathrm{DP}}$) is the detuning, and $\Omega_{\mathrm{SP}}$ ($\Omega_{\mathrm{DP}}$) is the Rabi frequency of the 468 (1079) nm light. (b) The $S_{1/2}$ to $P_{1/2}$ spectrum with $\Delta_{\mathrm{SP}}$ and $\Delta_{\mathrm{DP}}$ set to amplify ion motion. The local slope at $\Delta_{\mathrm{SP}}$ (red dashed line) is negative, while the global slope (blue dashed line) is positive. (c) OMS geometry with two \ce{Ra+} and one \ce{RaOCH3+} shown between two radial trap electrodes, as well as the relative orientation of the cooling and repump light and the magnetic field.}
    \label{fig:trap}
\end{figure}

\textit{Secular motion amplification by coherent population trapping}---We use CPT in \ce{Ra+} to amplify the ion crystal's motion. For Doppler cooling,  the cooling laser at 468 nm is red-detuned, $\Delta_{\mathrm{SP}} < 0$, from the $S_{1/2} \rightarrow P_{1/2}$ transition, and a repump laser at 1079 nm is blue-detuned, $\Delta_{\mathrm{DP}} > 0$,  from the $D_{3/2} \rightarrow P_{1/2}$ transition that brings the population back into the cooling cycle \cite{Berkeland2002}, see Fig. \ref{fig:trap}.  The high scattering rate of laser cooling can be significantly reduced by the CPT that occurs when $\Delta_{\mathrm{SP}} = \Delta_{\mathrm{DP}}$ \cite{Janik1985}. For CPT motional amplification
the condition is less stringent as setting $\Delta_{\mathrm{SP}} < \Delta_{\mathrm{DP}} < 0$ (see Fig. \ref{fig:trap}), heats the ion crystal because the 468 nm spectrum has a local slope that is negative due to the excitation suppression from CPT at $\Delta_{\mathrm{SP}} = \Delta_{\mathrm{DP}}$. However, the heating is bounded by the 468 nm global spectrum which has a positive slope for $\Delta_{\mathrm{SP}} < 0$, that cools the heated ion's motion once it reaches a sufficient amplitude where the global cooling spectrum is Doppler shifted into resonance. The trapped ions then maintain an amplified equilibrium orbit when the optically induced ``local heating'' and ``global cooling'' effects balance \cite{Supplemental, Bluemel1989}.

\begin{figure}
    \centering
    \includegraphics[width=\figsize\linewidth]{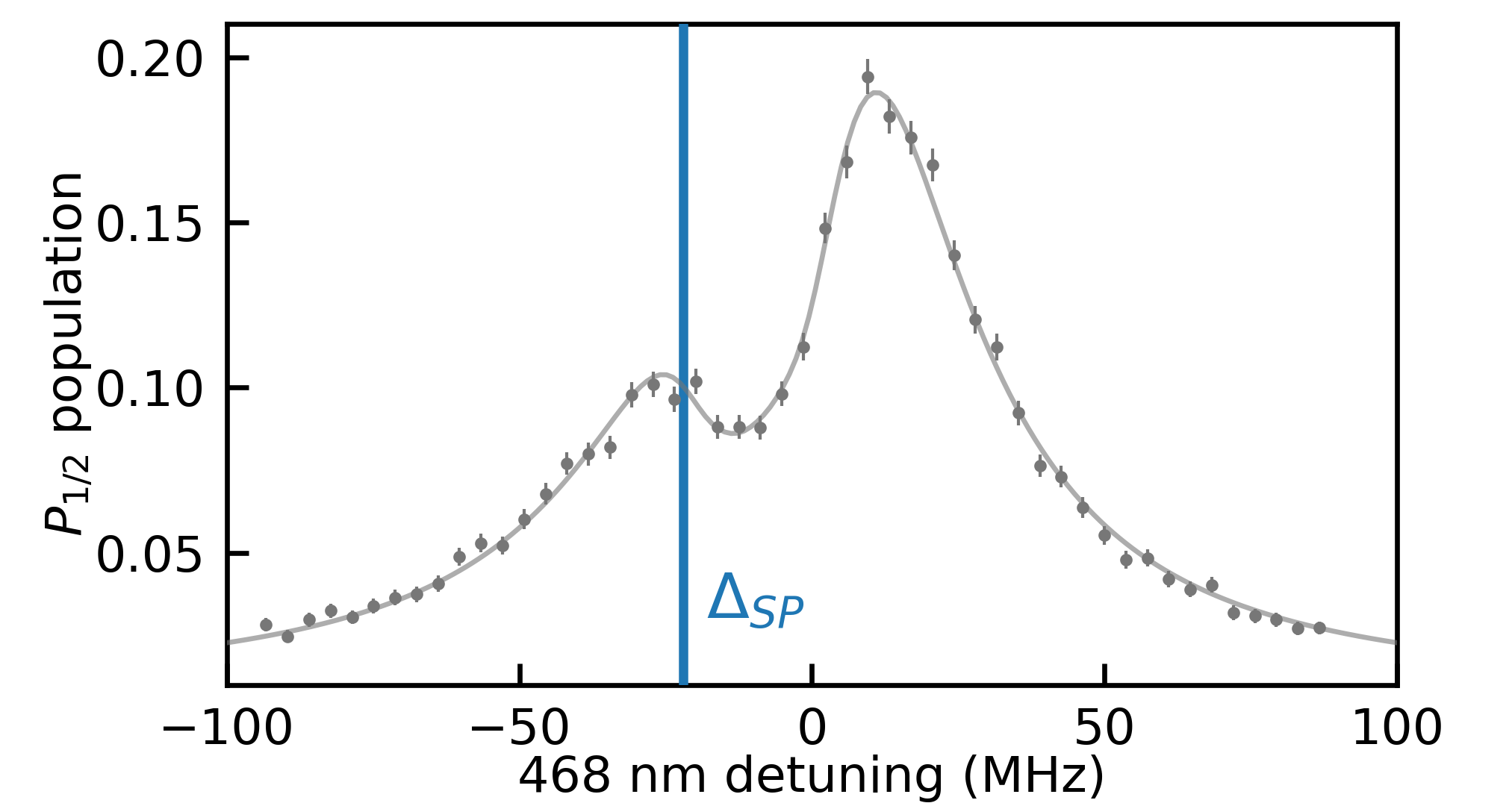}
    \caption{$P_{1/2}$ state population as a function of 468 nm detuning. A fit of the spectrum to a numerical solution of the $\Lambda$-level system that accounts for all Zeeman levels \cite{Oberst1999, Rossnagel2015} gives:
    $\Delta_{\mathrm{DP}}/2\pi=\SI{-10}{\mega\hertz}$, $\Omega_{\mathrm{SP}}/2\pi=\SI{19}{\mega\hertz}$, and $\Omega_{\mathrm{DP}}/2\pi=\SI{13}{\mega\hertz}$. The blue line at $\Delta_{\mathrm{SP}}/2\pi = \SI{-22}{\mega\hertz}$ is the detuning of the 468 nm light used for CPT amplification. The $P_{1/2}$ state population is not suppressed to zero at $\Delta_{\mathrm{SP}} = \Delta_{\mathrm{DP}}$ due to the finite linewidths of both the 468 and 1079 nm lasers ($\sim3$ MHz).}
    \label{fig:cpt}
\end{figure}

For OMS identification of \ce{RaOH+} and \ce{RaOCH3+}, we apply a 2.5 gauss magnetic field along the trap axial direction. The $k$ vectors of both lasers are at $45^{\circ}$ with respect to all trap axes and are linearly polarized perpendicular to the magnetic field direction (see Fig. \ref{fig:trap}). Each laser's frequency and amplitude is controlled with an acousto-optical modulator.
The multipeak spectrum of the cooling laser (see  Fig. \ref{fig:cpt}),  enables ``local heating, global cooling'' that amplifies the ion motion up to a fixed value. The 1079 nm light, with a $k$ vector perpendicular to the 468 nm light, see Fig. \ref{fig:trap} (c), breaks the degeneracy between the axial and radial directions so that the CPT only amplifies motion along the axial direction, see \cite{Supplemental}. To switch from CPT amplification to Doppler cooling, we detune $\Delta_{\mathrm{DP}}$ positive so the CPT excitation suppression is far from $\Delta_{\mathrm{SP}}$.

\textit{OMS of radium-based molecular ions}---We trap \ce{^{226}Ra+} ions in a linear Paul trap with a radio frequency (rf) drive of $\Omega_{rf}/2\pi=\SI{1}{\mega\hertz}$, radial electrode to trap center distance $r_0=\SI{3.0}{\milli\meter}$ and axial electrode to trap center distance $z_0=\SI{7.5}{\milli\meter}$ (for details see \cite{Fan2019}). For a single \ce{Ra+}, the axial secular frequency is $\omega_z/2\pi = \SI{27.8}{\kilo\hertz}$. The 468 nm fluorescent photons are collected and sent to a camera and a photomultiplier tube (PMT), and the counts are time-tagged using a field-programmable gate array \cite{Pruttivarasin2015}. 

\begin{figure}
    \centering
    \includegraphics[width=\figsize\linewidth]{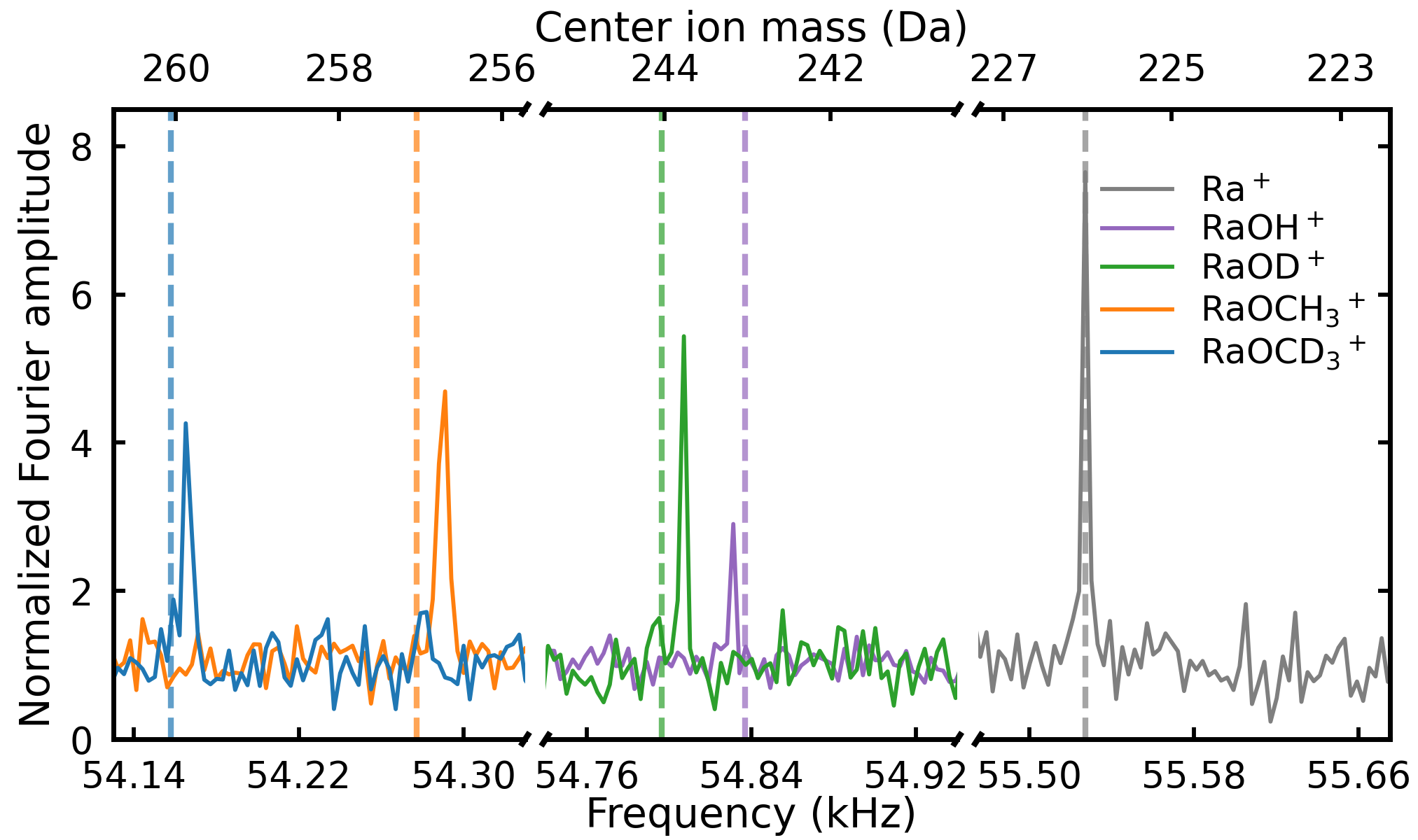}
    \caption{Fourier transformed PMT counts for Coulomb crystals where two \ce{Ra+} surround a third ion, labeled in the legend, in a linear chain. The Fourier amplitudes are normalized by their backgrounds for clarity. The dashed vertical lines show the calculated center ion masses. The peak amplitudes vary due to drift in the power and frequency of the 468 and 1079 nm lasers over the span of several days during which the measurements were taken. }
    \label{fig:ra}
\end{figure}

We laser cool three \ce{Ra+} ions and apply CPT amplification to increase the secular motion amplitude on the axial center-of-mass (COM) mode to $\SI{22\pm3}{\micro\meter}$.  A Fourier transform of the PMT counts while the ion motion is CPT amplified gives the OMS signal. The signal is calibrated by using known (fluorescing) ions. In this case three \ce{Ra+} ions are used for calibration (see Fig. \ref{fig:ra}). Next, either methanol vapor or the deuterated equivalent is introduced to react with the laser-cooled \ce{Ra+}. A chemical reaction produces a dark ion defect in the crystal and drops the PMT counts by roughly 1/3. If the dark ion is not in the middle of the crystal we re-order the ions to meet this condition by blue-detuning the 468 nm light, $\Delta_{\mathrm{SP}} > 0$, for $\sim\SI{1}{\second}$, which heats the trapped ions. We apply OMS to find the secular frequency and with the calibration measurement we can calculate the center ion's mass \cite{Supplemental}, see Fig. \ref{fig:ra}. Each trace is an average of ten $1/3$ s long measurements. The second harmonics of the secular frequencies are used because in our experimental setup they are the strongest Fourier components.

The difference in the mass spectrum of molecules when the trapped ions are exposed to methanol versus deuterated methanol confirms that we are producing the molecular ions identified by mass. When methanol is introduced only \ce{RaOCH3+} \cite{Lu1998, Lee2002a, Puri2017} and \ce{RaOH+} \cite{Augenbraun2020} are created, while \ce{RaOCD3+} and \ce{RaOD+} are only formed with deuterated methanol. The differences between the measured and calculated second harmonics of the secular frequencies are all within $\SI{13}{\hertz}$, corresponding to a fractional mass difference of $m / \Delta m \sim 800$ in a 3 s measurement. We observe that with a methanol (deuterated) background pressure of $\sim\SI{5e-10}{}$ torr, \ce{RaOH+} (\ce{RaOD+}) is not chemically stable and typically reacts in a few minutes to form \ce{RaOCH3+} (\ce{RaOCD3+}), which is easily detected because the OMS technique is fast, precise, and broadband.

\begin{table}[]
    \centering
    \begin{ruledtabular}
    \begin{tabular}{lcccc}
         & \ce{RaOH+} & \ce{RaOD+} & \ce{RaOCH3+} & \ce{RaOCD3+} \\
        \hline \\ [-0.6pc]
        Stat. & 243.19(7) & 243.79(7) & 256.72(8) & 259.85(8) \\
        Syst. & 0.01(11) & 0.21(11) & 0.34(11) & 0.22(11) \\
        Final & 243.20(14) & 244.01(14) & 257.06(14) & 260.07(14) \\
        [-0.7pc]\\
        Calc. & 243.03 & 244.03 & 257.04 & 260.06
    \end{tabular}
    \end{ruledtabular}
    \caption{Statistical results (Stat.) and systematic shifts and uncertainties (Syst.)  of the radium-based molecular ion masses measured by OMS in daltons.  The final molecular ion masses are calculated from a linear sum of the shifts, and the final uncertainties are given by summing the uncertainties in quadrature. See \cite{Supplemental} for details on the systematics. For comparison the calculated molecular ion masses (Calc.) are given \cite{Coursey2015}.}
    \label{tab:mass}
\end{table}

The OMS statistical uncertainty of 3 Hz was set by the Fourier transform resolution, which in turn comes from a $1/3$ $\SI{}{\second}$ data acquisition memory limit.  The line center is found with a Lorentzian fit, which has an uncertainty ($< 0.1$ Hz) that is much less than the Fourier frequency resolution. We also consider systematic effects including trap potential drift, secular motion amplitude shifts, and micromotion shifts \cite{Supplemental}. Both the statistical and systematic effects contribute to the ion mass uncertainty or shift by much less than 1 dalton, as summarized in Table \ref{tab:mass}.

\textit{OMS in a high frequency ion trap}---In a separate experimental apparatus, we confirm the OMS technique with strontium isotopes 88, 86, and 84, which we also identify with fluorescence. We demonstrate that the statistical mass sensitivity can be enhanced with a higher frequency ion trap ($r_0=\SI{0.6}{\milli\meter}$, $z_0=\SI{2.5}{\milli\meter}$, $\Omega_{\mathrm{rf}}/2\pi=\SI{22}{\mega\hertz}$). The axial secular frequency for a single \ce{^{88}Sr+} is $\omega_z/2\pi=\SI{91.7}{\kilo\hertz}$. 
We trap two-ion crystals with one \ce{^{88}Sr+} and one \ce{^{88, 87, 86, 84}Sr+}, and measure the axial COM secular frequencies of the crystal by OMS with CPT amplification applied to the \ce{^{88}Sr+}.
\ce{^{87}Sr+} is identified only with OMS.

\begin{figure}
    \centering
    \includegraphics[width=\figsize\linewidth]{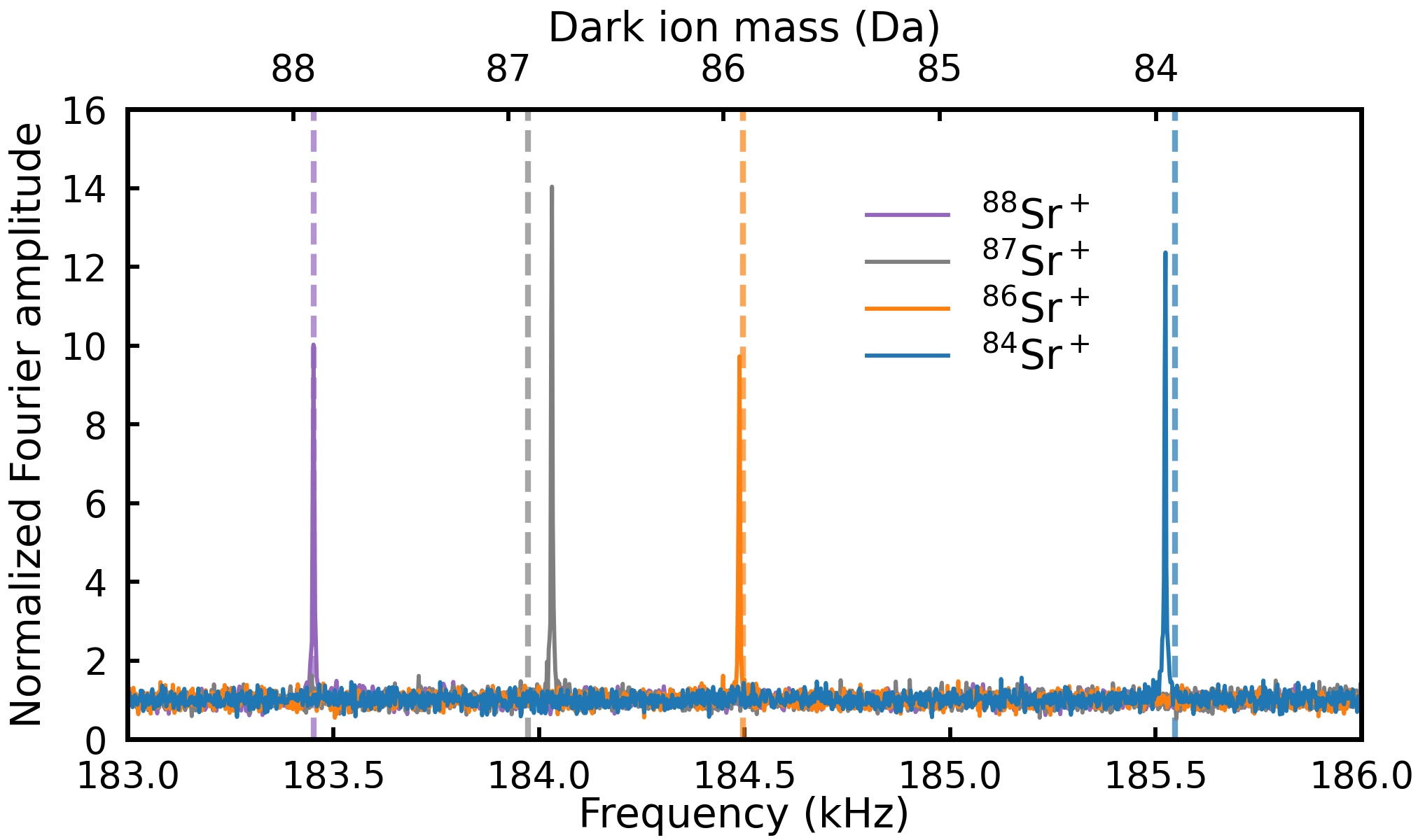}
    \caption{Fourier spectra of the fluorescent light from \ce{^{88}Sr+} which is in a crystal with a second \ce{^{88}Sr+}, \ce{^{87}Sr+}, \ce{^{86}Sr+}, or \ce{^{84}Sr+} ion. The calculated dark ion masses for the above crystals are shown as dashed vertical lines in the plot \cite{Supplemental}.}
    \label{fig:sr}
\end{figure}

The Fourier spectra of second harmonics of the axial COM secular frequency are shown in Fig. \ref{fig:sr}. The mass labels are calibrated using two $^{88}\ce{Sr+}$ ions \cite{Supplemental}. Each trace is an average of ten 0.5 s measurements. The corresponding statistical mass resolution is $m / \Delta m_{\mathrm{stat}} \sim \num{20000}$. The sub 1 dalton discrepancies between measured and calculated masses are primarily due to temporal drift of the trap potential.

\textit{Conclusion}---We have produced \ce{^{226}RaOH+} and \ce{^{226}RaOCH3+} molecules in this work. Their isotopologues \ce{^{225}RaOH+} and \ce{^{225}RaOCH3+} are proposed for nuclear Schiff moment measurements \cite{Kozyryev2017, Flambaum2019}, and can be produced with trapped \ce{^{225}Ra+} and methanol. The production, sympathetic cooling, and fast nondestructive identification of these radioactive polyatomic molecular ions enables studying their internal structure with techniques such as photodissociation spectroscopy \cite{Khanyile2015} or quantum logic spectroscopy \cite{Schmidt2005, Wolf2016}.

The short measurement time and sub-dalton mass resolution could enable detection of short-lived molecular ions, as demonstrated in this Letter using \ce{RaOH+} and \ce{RaOD+} which are metastable when in the presence of a methanol background. We note that the measurement time can be further reduced to $\SI{0.1}{\second}$ or less with an iris to selectively collect ion fluorescence on one end of the amplified ion motion. With such a short measurement time it is possible to apply this technique to study the dissociation channels of radium-based molecular ions. For example, \ce{Ra+} in the $D_{5/2}$ state (lifetime $\SI{0.3}{\second}$ \cite{Pal2009}) is optically indistinguishable from a radium-based molecular ion as neither of them fluoresce during \ce{Ra+} Doppler cooling, but can be distinguished with mass spectrometry. This provides a new tool to study the electronic states of chemical reaction products. The simultaneous determination of product and reactant states can be achieved by combining this method with optical pumping, which will help investigate full reaction pathways with single particles \cite{Sikorsky2018, Ratschbacher2012}.  The technique may also be used to rapidly characterize the motional frequencies of an ion trap. For example, it could be incorporated into a feedback loop for the long-term stabilization of trap motional frequencies \cite{Johnson2016}.

We thank W. Campbell, N. Hutzler and D. Patterson for helpful discussions. This research was performed under the sponsorship of the NSF Grant No.~PHY-1912665, and the University of California Office of the President (Grant No. MRP-19-601445).


\begin{thebibliography}{52}%
\makeatletter
\providecommand \@ifxundefined [1]{%
 \@ifx{#1\undefined}
}%
\providecommand \@ifnum [1]{%
 \ifnum #1\expandafter \@firstoftwo
 \else \expandafter \@secondoftwo
 \fi
}%
\providecommand \@ifx [1]{%
 \ifx #1\expandafter \@firstoftwo
 \else \expandafter \@secondoftwo
 \fi
}%
\providecommand \natexlab [1]{#1}%
\providecommand \enquote  [1]{``#1''}%
\providecommand \bibnamefont  [1]{#1}%
\providecommand \bibfnamefont [1]{#1}%
\providecommand \citenamefont [1]{#1}%
\providecommand \href@noop [0]{\@secondoftwo}%
\providecommand \href [0]{\begingroup \@sanitize@url \@href}%
\providecommand \@href[1]{\@@startlink{#1}\@@href}%
\providecommand \@@href[1]{\endgroup#1\@@endlink}%
\providecommand \@sanitize@url [0]{\catcode `\\12\catcode `\$12\catcode
  `\&12\catcode `\#12\catcode `\^12\catcode `\_12\catcode `\%12\relax}%
\providecommand \@@startlink[1]{}%
\providecommand \@@endlink[0]{}%
\providecommand \url  [0]{\begingroup\@sanitize@url \@url }%
\providecommand \@url [1]{\endgroup\@href {#1}{\urlprefix }}%
\providecommand \urlprefix  [0]{URL }%
\providecommand \Eprint [0]{\href }%
\providecommand \doibase [0]{https://doi.org/}%
\providecommand \selectlanguage [0]{\@gobble}%
\providecommand \bibinfo  [0]{\@secondoftwo}%
\providecommand \bibfield  [0]{\@secondoftwo}%
\providecommand \translation [1]{[#1]}%
\providecommand \BibitemOpen [0]{}%
\providecommand \bibitemStop [0]{}%
\providecommand \bibitemNoStop [0]{.\EOS\space}%
\providecommand \EOS [0]{\spacefactor3000\relax}%
\providecommand \BibitemShut  [1]{\csname bibitem#1\endcsname}%
\let\auto@bib@innerbib\@empty
\bibitem [{\citenamefont {Brewer}\ \emph {et~al.}(2019)\citenamefont {Brewer},
  \citenamefont {Chen}, \citenamefont {Hankin}, \citenamefont {Clements},
  \citenamefont {Chou}, \citenamefont {Wineland}, \citenamefont {Hume},\ and\
  \citenamefont {Leibrandt}}]{Brewer2019}%
  \BibitemOpen
  \bibfield  {author} {\bibinfo {author} {\bibfnamefont {S.~M.}\ \bibnamefont
  {Brewer}}, \bibinfo {author} {\bibfnamefont {J.-S.}\ \bibnamefont {Chen}},
  \bibinfo {author} {\bibfnamefont {A.~M.}\ \bibnamefont {Hankin}}, \bibinfo
  {author} {\bibfnamefont {E.~R.}\ \bibnamefont {Clements}}, \bibinfo {author}
  {\bibfnamefont {C.~W.}\ \bibnamefont {Chou}}, \bibinfo {author}
  {\bibfnamefont {D.~J.}\ \bibnamefont {Wineland}}, \bibinfo {author}
  {\bibfnamefont {D.~B.}\ \bibnamefont {Hume}},\ and\ \bibinfo {author}
  {\bibfnamefont {D.~R.}\ \bibnamefont {Leibrandt}},\ }\href
  {https://link.aps.org/doi/10.1103/PhysRevLett.123.033201} {\bibfield
  {journal} {\bibinfo  {journal} {Phys. Rev. Lett.}\ }\textbf {\bibinfo
  {volume} {123}},\ \bibinfo {pages} {033201} (\bibinfo {year}
  {2019})}\BibitemShut {NoStop}%
\bibitem [{\citenamefont {Cairncross}\ \emph {et~al.}(2017)\citenamefont
  {Cairncross}, \citenamefont {Gresh}, \citenamefont {Grau}, \citenamefont
  {Cossel}, \citenamefont {Roussy}, \citenamefont {Ni}, \citenamefont {Zhou},
  \citenamefont {Ye},\ and\ \citenamefont {Cornell}}]{Cairncross2017}%
  \BibitemOpen
  \bibfield  {author} {\bibinfo {author} {\bibfnamefont {W.~B.}\ \bibnamefont
  {Cairncross}}, \bibinfo {author} {\bibfnamefont {D.~N.}\ \bibnamefont
  {Gresh}}, \bibinfo {author} {\bibfnamefont {M.}~\bibnamefont {Grau}},
  \bibinfo {author} {\bibfnamefont {K.~C.}\ \bibnamefont {Cossel}}, \bibinfo
  {author} {\bibfnamefont {T.~S.}\ \bibnamefont {Roussy}}, \bibinfo {author}
  {\bibfnamefont {Y.}~\bibnamefont {Ni}}, \bibinfo {author} {\bibfnamefont
  {Y.}~\bibnamefont {Zhou}}, \bibinfo {author} {\bibfnamefont {J.}~\bibnamefont
  {Ye}},\ and\ \bibinfo {author} {\bibfnamefont {E.~A.}\ \bibnamefont
  {Cornell}},\ }\href {https://link.aps.org/doi/10.1103/PhysRevLett.119.153001}
  {\bibfield  {journal} {\bibinfo  {journal} {Phys. Rev. Lett.}\ }\textbf
  {\bibinfo {volume} {119}},\ \bibinfo {pages} {153001} (\bibinfo {year}
  {2017})}\BibitemShut {NoStop}%
\bibitem [{\citenamefont {Puri}\ \emph {et~al.}(2019)\citenamefont {Puri},
  \citenamefont {Mills}, \citenamefont {Simbotin}, \citenamefont {Montgomery},
  \citenamefont {Côté}, \citenamefont {Schneider}, \citenamefont {Suits},\
  and\ \citenamefont {Hudson}}]{Puri2019}%
  \BibitemOpen
  \bibfield  {author} {\bibinfo {author} {\bibfnamefont {P.}~\bibnamefont
  {Puri}}, \bibinfo {author} {\bibfnamefont {M.}~\bibnamefont {Mills}},
  \bibinfo {author} {\bibfnamefont {I.}~\bibnamefont {Simbotin}}, \bibinfo
  {author} {\bibfnamefont {J.~A.}\ \bibnamefont {Montgomery}}, \bibinfo
  {author} {\bibfnamefont {R.}~\bibnamefont {Côté}}, \bibinfo {author}
  {\bibfnamefont {C.}~\bibnamefont {Schneider}}, \bibinfo {author}
  {\bibfnamefont {A.~G.}\ \bibnamefont {Suits}},\ and\ \bibinfo {author}
  {\bibfnamefont {E.~R.}\ \bibnamefont {Hudson}},\ }\href
  {https://doi.org/10.1038/s41557-019-0264-3} {\bibfield  {journal} {\bibinfo
  {journal} {Nat. Chem.}\ }\textbf {\bibinfo {volume} {11}},\ \bibinfo {pages}
  {615} (\bibinfo {year} {2019})}\BibitemShut {NoStop}%
\bibitem [{\citenamefont {Sikorsky}\ \emph {et~al.}(2018)\citenamefont
  {Sikorsky}, \citenamefont {Meir}, \citenamefont {Ben-shlomi}, \citenamefont
  {Akerman},\ and\ \citenamefont {Ozeri}}]{Sikorsky2018}%
  \BibitemOpen
  \bibfield  {author} {\bibinfo {author} {\bibfnamefont {T.}~\bibnamefont
  {Sikorsky}}, \bibinfo {author} {\bibfnamefont {Z.}~\bibnamefont {Meir}},
  \bibinfo {author} {\bibfnamefont {R.}~\bibnamefont {Ben-shlomi}}, \bibinfo
  {author} {\bibfnamefont {N.}~\bibnamefont {Akerman}},\ and\ \bibinfo {author}
  {\bibfnamefont {R.}~\bibnamefont {Ozeri}},\ }\href
  {https://doi.org/10.1038/s41467-018-03373-y} {\bibfield  {journal} {\bibinfo
  {journal} {Nature Communications}\ }\textbf {\bibinfo {volume} {9}},\
  \bibinfo {pages} {920} (\bibinfo {year} {2018})}\BibitemShut {NoStop}%
\bibitem [{\citenamefont {Safronova}\ \emph {et~al.}(2014)\citenamefont
  {Safronova}, \citenamefont {Dzuba}, \citenamefont {Flambaum}, \citenamefont
  {Safronova}, \citenamefont {Porsev},\ and\ \citenamefont
  {Kozlov}}]{Safronova2014}%
  \BibitemOpen
  \bibfield  {author} {\bibinfo {author} {\bibfnamefont {M.~S.}\ \bibnamefont
  {Safronova}}, \bibinfo {author} {\bibfnamefont {V.~A.}\ \bibnamefont
  {Dzuba}}, \bibinfo {author} {\bibfnamefont {V.~V.}\ \bibnamefont {Flambaum}},
  \bibinfo {author} {\bibfnamefont {U.~I.}\ \bibnamefont {Safronova}}, \bibinfo
  {author} {\bibfnamefont {S.~G.}\ \bibnamefont {Porsev}},\ and\ \bibinfo
  {author} {\bibfnamefont {M.~G.}\ \bibnamefont {Kozlov}},\ }\href
  {https://link.aps.org/doi/10.1103/PhysRevLett.113.030801} {\bibfield
  {journal} {\bibinfo  {journal} {Phys. Rev. Lett.}\ }\textbf {\bibinfo
  {volume} {113}},\ \bibinfo {pages} {030801} (\bibinfo {year}
  {2014})}\BibitemShut {NoStop}%
\bibitem [{\citenamefont {Kozlov}\ \emph {et~al.}(2018)\citenamefont {Kozlov},
  \citenamefont {Safronova}, \citenamefont {Crespo López-Urrutia},\ and\
  \citenamefont {Schmidt}}]{Kozlov2018}%
  \BibitemOpen
  \bibfield  {author} {\bibinfo {author} {\bibfnamefont {M.~G.}\ \bibnamefont
  {Kozlov}}, \bibinfo {author} {\bibfnamefont {M.~S.}\ \bibnamefont
  {Safronova}}, \bibinfo {author} {\bibfnamefont {J.~R.}\ \bibnamefont {Crespo
  López-Urrutia}},\ and\ \bibinfo {author} {\bibfnamefont {P.~O.}\
  \bibnamefont {Schmidt}},\ }\href
  {https://link.aps.org/doi/10.1103/RevModPhys.90.045005} {\bibfield  {journal}
  {\bibinfo  {journal} {Rev. Mod. Phys.}\ }\textbf {\bibinfo {volume} {90}},\
  \bibinfo {pages} {045005} (\bibinfo {year} {2018})}\BibitemShut {NoStop}%
\bibitem [{\citenamefont {Micke}\ \emph {et~al.}(2020)\citenamefont {Micke},
  \citenamefont {Leopold}, \citenamefont {King}, \citenamefont {Benkler},
  \citenamefont {Spieß}, \citenamefont {Schmöger}, \citenamefont {Schwarz},
  \citenamefont {Crespo López-Urrutia},\ and\ \citenamefont
  {Schmidt}}]{Micke2020}%
  \BibitemOpen
  \bibfield  {author} {\bibinfo {author} {\bibfnamefont {P.}~\bibnamefont
  {Micke}}, \bibinfo {author} {\bibfnamefont {T.}~\bibnamefont {Leopold}},
  \bibinfo {author} {\bibfnamefont {S.~A.}\ \bibnamefont {King}}, \bibinfo
  {author} {\bibfnamefont {E.}~\bibnamefont {Benkler}}, \bibinfo {author}
  {\bibfnamefont {L.~J.}\ \bibnamefont {Spieß}}, \bibinfo {author}
  {\bibfnamefont {L.}~\bibnamefont {Schmöger}}, \bibinfo {author}
  {\bibfnamefont {M.}~\bibnamefont {Schwarz}}, \bibinfo {author} {\bibfnamefont
  {J.~R.}\ \bibnamefont {Crespo López-Urrutia}},\ and\ \bibinfo {author}
  {\bibfnamefont {P.~O.}\ \bibnamefont {Schmidt}},\ }\href
  {https://doi.org/10.1038/s41586-020-1959-8} {\bibfield  {journal} {\bibinfo
  {journal} {Nature}\ }\textbf {\bibinfo {volume} {578}},\ \bibinfo {pages}
  {60} (\bibinfo {year} {2020})}\BibitemShut {NoStop}%
\bibitem [{\citenamefont {Andreev}\ \emph {et~al.}(2018)\citenamefont
  {Andreev}, \citenamefont {Ang}, \citenamefont {DeMille}, \citenamefont
  {Doyle}, \citenamefont {Gabrielse}, \citenamefont {Haefner}, \citenamefont
  {Hutzler}, \citenamefont {Lasner}, \citenamefont {Meisenhelder},
  \citenamefont {O’Leary}, \citenamefont {Panda}, \citenamefont {West},
  \citenamefont {West}, \citenamefont {Wu},\ and\ \citenamefont
  {Collaboration}}]{Andreev2018}%
  \BibitemOpen
  \bibfield  {author} {\bibinfo {author} {\bibfnamefont {V.}~\bibnamefont
  {Andreev}}, \bibinfo {author} {\bibfnamefont {D.~G.}\ \bibnamefont {Ang}},
  \bibinfo {author} {\bibfnamefont {D.}~\bibnamefont {DeMille}}, \bibinfo
  {author} {\bibfnamefont {J.~M.}\ \bibnamefont {Doyle}}, \bibinfo {author}
  {\bibfnamefont {G.}~\bibnamefont {Gabrielse}}, \bibinfo {author}
  {\bibfnamefont {J.}~\bibnamefont {Haefner}}, \bibinfo {author} {\bibfnamefont
  {N.~R.}\ \bibnamefont {Hutzler}}, \bibinfo {author} {\bibfnamefont
  {Z.}~\bibnamefont {Lasner}}, \bibinfo {author} {\bibfnamefont
  {C.}~\bibnamefont {Meisenhelder}}, \bibinfo {author} {\bibfnamefont {B.~R.}\
  \bibnamefont {O’Leary}}, \bibinfo {author} {\bibfnamefont {C.~D.}\
  \bibnamefont {Panda}}, \bibinfo {author} {\bibfnamefont {A.~D.}\ \bibnamefont
  {West}}, \bibinfo {author} {\bibfnamefont {E.~P.}\ \bibnamefont {West}},
  \bibinfo {author} {\bibfnamefont {X.}~\bibnamefont {Wu}},\ and\ \bibinfo
  {author} {\bibfnamefont {A.~C. M.~E.}\ \bibnamefont {Collaboration}},\ }\href
  {https://doi.org/10.1038/s41586-018-0599-8} {\bibfield  {journal} {\bibinfo
  {journal} {Nature}\ }\textbf {\bibinfo {volume} {562}},\ \bibinfo {pages}
  {355} (\bibinfo {year} {2018})}\BibitemShut {NoStop}%
\bibitem [{\citenamefont {Flambaum}(2019)}]{Flambaum2019}%
  \BibitemOpen
  \bibfield  {author} {\bibinfo {author} {\bibfnamefont {V.~V.}\ \bibnamefont
  {Flambaum}},\ }\href {https://link.aps.org/doi/10.1103/PhysRevC.99.035501}
  {\bibfield  {journal} {\bibinfo  {journal} {Phys. Rev. C}\ }\textbf {\bibinfo
  {volume} {99}},\ \bibinfo {pages} {035501} (\bibinfo {year}
  {2019})}\BibitemShut {NoStop}%
\bibitem [{\citenamefont {Kozyryev}\ and\ \citenamefont
  {Hutzler}(2017)}]{Kozyryev2017}%
  \BibitemOpen
  \bibfield  {author} {\bibinfo {author} {\bibfnamefont {I.}~\bibnamefont
  {Kozyryev}}\ and\ \bibinfo {author} {\bibfnamefont {N.~R.}\ \bibnamefont
  {Hutzler}},\ }\href {https://link.aps.org/doi/10.1103/PhysRevLett.119.133002}
  {\bibfield  {journal} {\bibinfo  {journal} {Phys. Rev. Lett.}\ }\textbf
  {\bibinfo {volume} {119}},\ \bibinfo {pages} {133002} (\bibinfo {year}
  {2017})}\BibitemShut {NoStop}%
\bibitem [{\citenamefont {Garcia~Ruiz}\ \emph {et~al.}(2020)\citenamefont
  {Garcia~Ruiz}, \citenamefont {Berger}, \citenamefont {Billowes},
  \citenamefont {Binnersley}, \citenamefont {Bissell}, \citenamefont {Breier},
  \citenamefont {Brinson}, \citenamefont {Chrysalidis}, \citenamefont
  {Cocolios}, \citenamefont {Cooper}, \citenamefont {Flanagan}, \citenamefont
  {Giesen}, \citenamefont {de~Groote}, \citenamefont {Franchoo}, \citenamefont
  {Gustafsson}, \citenamefont {Isaev}, \citenamefont {Koszorús}, \citenamefont
  {Neyens}, \citenamefont {Perrett}, \citenamefont {Ricketts}, \citenamefont
  {Rothe}, \citenamefont {Schweikhard}, \citenamefont {Vernon}, \citenamefont
  {Wendt}, \citenamefont {Wienholtz}, \citenamefont {Wilkins},\ and\
  \citenamefont {Yang}}]{GarciaRuiz2020}%
  \BibitemOpen
  \bibfield  {author} {\bibinfo {author} {\bibfnamefont {R.~F.}\ \bibnamefont
  {Garcia~Ruiz}}, \bibinfo {author} {\bibfnamefont {R.}~\bibnamefont {Berger}},
  \bibinfo {author} {\bibfnamefont {J.}~\bibnamefont {Billowes}}, \bibinfo
  {author} {\bibfnamefont {C.~L.}\ \bibnamefont {Binnersley}}, \bibinfo
  {author} {\bibfnamefont {M.~L.}\ \bibnamefont {Bissell}}, \bibinfo {author}
  {\bibfnamefont {A.~A.}\ \bibnamefont {Breier}}, \bibinfo {author}
  {\bibfnamefont {A.~J.}\ \bibnamefont {Brinson}}, \bibinfo {author}
  {\bibfnamefont {K.}~\bibnamefont {Chrysalidis}}, \bibinfo {author}
  {\bibfnamefont {T.~E.}\ \bibnamefont {Cocolios}}, \bibinfo {author}
  {\bibfnamefont {B.~S.}\ \bibnamefont {Cooper}}, \bibinfo {author}
  {\bibfnamefont {K.~T.}\ \bibnamefont {Flanagan}}, \bibinfo {author}
  {\bibfnamefont {T.~F.}\ \bibnamefont {Giesen}}, \bibinfo {author}
  {\bibfnamefont {R.~P.}\ \bibnamefont {de~Groote}}, \bibinfo {author}
  {\bibfnamefont {S.}~\bibnamefont {Franchoo}}, \bibinfo {author}
  {\bibfnamefont {F.~P.}\ \bibnamefont {Gustafsson}}, \bibinfo {author}
  {\bibfnamefont {T.~A.}\ \bibnamefont {Isaev}}, \bibinfo {author}
  {\bibfnamefont {{\'A}.}~\bibnamefont {Koszorús}}, \bibinfo {author}
  {\bibfnamefont {G.}~\bibnamefont {Neyens}}, \bibinfo {author} {\bibfnamefont
  {H.~A.}\ \bibnamefont {Perrett}}, \bibinfo {author} {\bibfnamefont {C.~M.}\
  \bibnamefont {Ricketts}}, \bibinfo {author} {\bibfnamefont {S.}~\bibnamefont
  {Rothe}}, \bibinfo {author} {\bibfnamefont {L.}~\bibnamefont {Schweikhard}},
  \bibinfo {author} {\bibfnamefont {A.~R.}\ \bibnamefont {Vernon}}, \bibinfo
  {author} {\bibfnamefont {K.~D.~A.}\ \bibnamefont {Wendt}}, \bibinfo {author}
  {\bibfnamefont {F.}~\bibnamefont {Wienholtz}}, \bibinfo {author}
  {\bibfnamefont {S.~G.}\ \bibnamefont {Wilkins}},\ and\ \bibinfo {author}
  {\bibfnamefont {X.~F.}\ \bibnamefont {Yang}},\ }\href
  {https://doi.org/10.1038/s41586-020-2299-4} {\bibfield  {journal} {\bibinfo
  {journal} {Nature}\ }\textbf {\bibinfo {volume} {581}},\ \bibinfo {pages}
  {396} (\bibinfo {year} {2020})}\BibitemShut {NoStop}%
\bibitem [{\citenamefont {Gaffney}\ \emph {et~al.}(2013)\citenamefont
  {Gaffney}, \citenamefont {Butler}, \citenamefont {Scheck}, \citenamefont
  {Hayes}, \citenamefont {Wenander}, \citenamefont {Albers}, \citenamefont
  {Bastin}, \citenamefont {Bauer}, \citenamefont {Blazhev}, \citenamefont
  {Bonig}, \citenamefont {Bree}, \citenamefont {Cederkall}, \citenamefont
  {Chupp}, \citenamefont {Cline}, \citenamefont {Cocolios}, \citenamefont
  {Davinson}, \citenamefont {De~Witte}, \citenamefont {Diriken}, \citenamefont
  {Grahn}, \citenamefont {Herzan}, \citenamefont {Huyse}, \citenamefont
  {Jenkins}, \citenamefont {Joss}, \citenamefont {Kesteloot}, \citenamefont
  {Konki}, \citenamefont {Kowalczyk}, \citenamefont {Kroll}, \citenamefont
  {Kwan}, \citenamefont {Lutter}, \citenamefont {Moschner}, \citenamefont
  {Napiorkowski}, \citenamefont {Pakarinen}, \citenamefont {Pfeiffer},
  \citenamefont {Radeck}, \citenamefont {Reiter}, \citenamefont {Reynders},
  \citenamefont {Rigby}, \citenamefont {Robledo}, \citenamefont {Rudigier},
  \citenamefont {Sambi}, \citenamefont {Seidlitz}, \citenamefont {Siebeck},
  \citenamefont {Stora}, \citenamefont {Thoele}, \citenamefont {Van~Duppen},
  \citenamefont {Vermeulen}, \citenamefont {von Schmid}, \citenamefont
  {Voulot}, \citenamefont {Warr}, \citenamefont {Wimmer}, \citenamefont
  {Wrzosek-Lipska}, \citenamefont {Wu},\ and\ \citenamefont
  {Zielinska}}]{Gaffney2013}%
  \BibitemOpen
  \bibfield  {author} {\bibinfo {author} {\bibfnamefont {L.~P.}\ \bibnamefont
  {Gaffney}}, \bibinfo {author} {\bibfnamefont {P.~A.}\ \bibnamefont {Butler}},
  \bibinfo {author} {\bibfnamefont {M.}~\bibnamefont {Scheck}}, \bibinfo
  {author} {\bibfnamefont {A.~B.}\ \bibnamefont {Hayes}}, \bibinfo {author}
  {\bibfnamefont {F.}~\bibnamefont {Wenander}}, \bibinfo {author}
  {\bibfnamefont {M.}~\bibnamefont {Albers}}, \bibinfo {author} {\bibfnamefont
  {B.}~\bibnamefont {Bastin}}, \bibinfo {author} {\bibfnamefont
  {C.}~\bibnamefont {Bauer}}, \bibinfo {author} {\bibfnamefont
  {A.}~\bibnamefont {Blazhev}}, \bibinfo {author} {\bibfnamefont
  {S.}~\bibnamefont {Bonig}}, \bibinfo {author} {\bibfnamefont
  {N.}~\bibnamefont {Bree}}, \bibinfo {author} {\bibfnamefont {J.}~\bibnamefont
  {Cederkall}}, \bibinfo {author} {\bibfnamefont {T.}~\bibnamefont {Chupp}},
  \bibinfo {author} {\bibfnamefont {D.}~\bibnamefont {Cline}}, \bibinfo
  {author} {\bibfnamefont {T.~E.}\ \bibnamefont {Cocolios}}, \bibinfo {author}
  {\bibfnamefont {T.}~\bibnamefont {Davinson}}, \bibinfo {author}
  {\bibfnamefont {H.}~\bibnamefont {De~Witte}}, \bibinfo {author}
  {\bibfnamefont {J.}~\bibnamefont {Diriken}}, \bibinfo {author} {\bibfnamefont
  {T.}~\bibnamefont {Grahn}}, \bibinfo {author} {\bibfnamefont
  {A.}~\bibnamefont {Herzan}}, \bibinfo {author} {\bibfnamefont
  {M.}~\bibnamefont {Huyse}}, \bibinfo {author} {\bibfnamefont {D.~G.}\
  \bibnamefont {Jenkins}}, \bibinfo {author} {\bibfnamefont {D.~T.}\
  \bibnamefont {Joss}}, \bibinfo {author} {\bibfnamefont {N.}~\bibnamefont
  {Kesteloot}}, \bibinfo {author} {\bibfnamefont {J.}~\bibnamefont {Konki}},
  \bibinfo {author} {\bibfnamefont {M.}~\bibnamefont {Kowalczyk}}, \bibinfo
  {author} {\bibfnamefont {T.}~\bibnamefont {Kroll}}, \bibinfo {author}
  {\bibfnamefont {E.}~\bibnamefont {Kwan}}, \bibinfo {author} {\bibfnamefont
  {R.}~\bibnamefont {Lutter}}, \bibinfo {author} {\bibfnamefont
  {K.}~\bibnamefont {Moschner}}, \bibinfo {author} {\bibfnamefont
  {P.}~\bibnamefont {Napiorkowski}}, \bibinfo {author} {\bibfnamefont
  {J.}~\bibnamefont {Pakarinen}}, \bibinfo {author} {\bibfnamefont
  {M.}~\bibnamefont {Pfeiffer}}, \bibinfo {author} {\bibfnamefont
  {D.}~\bibnamefont {Radeck}}, \bibinfo {author} {\bibfnamefont
  {P.}~\bibnamefont {Reiter}}, \bibinfo {author} {\bibfnamefont
  {K.}~\bibnamefont {Reynders}}, \bibinfo {author} {\bibfnamefont {S.~V.}\
  \bibnamefont {Rigby}}, \bibinfo {author} {\bibfnamefont {L.~M.}\ \bibnamefont
  {Robledo}}, \bibinfo {author} {\bibfnamefont {M.}~\bibnamefont {Rudigier}},
  \bibinfo {author} {\bibfnamefont {S.}~\bibnamefont {Sambi}}, \bibinfo
  {author} {\bibfnamefont {M.}~\bibnamefont {Seidlitz}}, \bibinfo {author}
  {\bibfnamefont {B.}~\bibnamefont {Siebeck}}, \bibinfo {author} {\bibfnamefont
  {T.}~\bibnamefont {Stora}}, \bibinfo {author} {\bibfnamefont
  {P.}~\bibnamefont {Thoele}}, \bibinfo {author} {\bibfnamefont
  {P.}~\bibnamefont {Van~Duppen}}, \bibinfo {author} {\bibfnamefont {M.~J.}\
  \bibnamefont {Vermeulen}}, \bibinfo {author} {\bibfnamefont {M.}~\bibnamefont
  {von Schmid}}, \bibinfo {author} {\bibfnamefont {D.}~\bibnamefont {Voulot}},
  \bibinfo {author} {\bibfnamefont {N.}~\bibnamefont {Warr}}, \bibinfo {author}
  {\bibfnamefont {K.}~\bibnamefont {Wimmer}}, \bibinfo {author} {\bibfnamefont
  {K.}~\bibnamefont {Wrzosek-Lipska}}, \bibinfo {author} {\bibfnamefont
  {C.~Y.}\ \bibnamefont {Wu}},\ and\ \bibinfo {author} {\bibfnamefont
  {M.}~\bibnamefont {Zielinska}},\ }\href
  {http://dx.doi.org/10.1038/nature12073} {\bibfield  {journal} {\bibinfo
  {journal} {Nature}\ }\textbf {\bibinfo {volume} {497}},\ \bibinfo {pages}
  {199} (\bibinfo {year} {2013})}\BibitemShut {NoStop}%
\bibitem [{\citenamefont {Graner}\ \emph {et~al.}(2016)\citenamefont {Graner},
  \citenamefont {Chen}, \citenamefont {Lindahl},\ and\ \citenamefont
  {Heckel}}]{Graner2016}%
  \BibitemOpen
  \bibfield  {author} {\bibinfo {author} {\bibfnamefont {B.}~\bibnamefont
  {Graner}}, \bibinfo {author} {\bibfnamefont {Y.}~\bibnamefont {Chen}},
  \bibinfo {author} {\bibfnamefont {E.~G.}\ \bibnamefont {Lindahl}},\ and\
  \bibinfo {author} {\bibfnamefont {B.~R.}\ \bibnamefont {Heckel}},\ }\href
  {https://link.aps.org/doi/10.1103/PhysRevLett.116.161601} {\bibfield
  {journal} {\bibinfo  {journal} {Phys. Rev. Lett.}\ }\textbf {\bibinfo
  {volume} {116}},\ \bibinfo {pages} {161601} (\bibinfo {year}
  {2016})}\BibitemShut {NoStop}%
\bibitem [{\citenamefont {Kudashov}\ \emph {et~al.}(2014)\citenamefont
  {Kudashov}, \citenamefont {Petrov}, \citenamefont {Skripnikov}, \citenamefont
  {Mosyagin}, \citenamefont {Isaev}, \citenamefont {Berger},\ and\
  \citenamefont {Titov}}]{Kudashov2014}%
  \BibitemOpen
  \bibfield  {author} {\bibinfo {author} {\bibfnamefont {A.~D.}\ \bibnamefont
  {Kudashov}}, \bibinfo {author} {\bibfnamefont {A.~N.}\ \bibnamefont
  {Petrov}}, \bibinfo {author} {\bibfnamefont {L.~V.}\ \bibnamefont
  {Skripnikov}}, \bibinfo {author} {\bibfnamefont {N.~S.}\ \bibnamefont
  {Mosyagin}}, \bibinfo {author} {\bibfnamefont {T.~A.}\ \bibnamefont {Isaev}},
  \bibinfo {author} {\bibfnamefont {R.}~\bibnamefont {Berger}},\ and\ \bibinfo
  {author} {\bibfnamefont {A.~V.}\ \bibnamefont {Titov}},\ }\href
  {https://link.aps.org/doi/10.1103/PhysRevA.90.052513} {\bibfield  {journal}
  {\bibinfo  {journal} {Phys. Rev. A}\ }\textbf {\bibinfo {volume} {90}},\
  \bibinfo {pages} {052513} (\bibinfo {year} {2014})}\BibitemShut {NoStop}%
\bibitem [{\citenamefont {Zhou}\ \emph {et~al.}(2020)\citenamefont {Zhou},
  \citenamefont {Shagam}, \citenamefont {Cairncross}, \citenamefont {Ng},
  \citenamefont {Roussy}, \citenamefont {Grogan}, \citenamefont {Boyce},
  \citenamefont {Vigil}, \citenamefont {Pettine}, \citenamefont {Zelevinsky},
  \citenamefont {Ye},\ and\ \citenamefont {Cornell}}]{Zhou2020}%
  \BibitemOpen
  \bibfield  {author} {\bibinfo {author} {\bibfnamefont {Y.}~\bibnamefont
  {Zhou}}, \bibinfo {author} {\bibfnamefont {Y.}~\bibnamefont {Shagam}},
  \bibinfo {author} {\bibfnamefont {W.~B.}\ \bibnamefont {Cairncross}},
  \bibinfo {author} {\bibfnamefont {K.~B.}\ \bibnamefont {Ng}}, \bibinfo
  {author} {\bibfnamefont {T.~S.}\ \bibnamefont {Roussy}}, \bibinfo {author}
  {\bibfnamefont {T.}~\bibnamefont {Grogan}}, \bibinfo {author} {\bibfnamefont
  {K.}~\bibnamefont {Boyce}}, \bibinfo {author} {\bibfnamefont
  {A.}~\bibnamefont {Vigil}}, \bibinfo {author} {\bibfnamefont
  {M.}~\bibnamefont {Pettine}}, \bibinfo {author} {\bibfnamefont
  {T.}~\bibnamefont {Zelevinsky}}, \bibinfo {author} {\bibfnamefont
  {J.}~\bibnamefont {Ye}},\ and\ \bibinfo {author} {\bibfnamefont {E.~A.}\
  \bibnamefont {Cornell}},\ }\href
  {https://doi.org/10.1103/PhysRevLett.124.053201} {\bibfield  {journal}
  {\bibinfo  {journal} {Phys. Rev. Lett.}\ }\textbf {\bibinfo {volume} {124}},\
  \bibinfo {pages} {053201} (\bibinfo {year} {2020})}\BibitemShut {NoStop}%
\bibitem [{\citenamefont {Yu}\ and\ \citenamefont {Hutzler}(2021)}]{Yu2021}%
  \BibitemOpen
  \bibfield  {author} {\bibinfo {author} {\bibfnamefont {P.}~\bibnamefont
  {Yu}}\ and\ \bibinfo {author} {\bibfnamefont {N.~R.}\ \bibnamefont
  {Hutzler}},\ }\href {https://doi.org/10.1103/PhysRevLett.126.023003}
  {\bibfield  {journal} {\bibinfo  {journal} {Phys. Rev. Lett.}\ }\textbf
  {\bibinfo {volume} {126}},\ \bibinfo {pages} {023003} (\bibinfo {year}
  {2021})}\BibitemShut {NoStop}%
\bibitem [{\citenamefont {Chen}\ \emph {et~al.}(2015)\citenamefont {Chen},
  \citenamefont {Lin},\ and\ \citenamefont {Odom}}]{Chen2015}%
  \BibitemOpen
  \bibfield  {author} {\bibinfo {author} {\bibfnamefont {X.}~\bibnamefont
  {Chen}}, \bibinfo {author} {\bibfnamefont {Y.-W.}\ \bibnamefont {Lin}},\ and\
  \bibinfo {author} {\bibfnamefont {B.~C.}\ \bibnamefont {Odom}},\ }\href
  {https://doi.org/10.1088/1367-2630/17/4/043037} {\bibfield  {journal}
  {\bibinfo  {journal} {New J. Phys.}\ }\textbf {\bibinfo {volume} {17}},\
  \bibinfo {pages} {043037} (\bibinfo {year} {2015})}\BibitemShut {NoStop}%
\bibitem [{\citenamefont {Berkeland}\ and\ \citenamefont
  {Boshier}(2002)}]{Berkeland2002}%
  \BibitemOpen
  \bibfield  {author} {\bibinfo {author} {\bibfnamefont {D.~J.}\ \bibnamefont
  {Berkeland}}\ and\ \bibinfo {author} {\bibfnamefont {M.~G.}\ \bibnamefont
  {Boshier}},\ }\href {https://link.aps.org/doi/10.1103/PhysRevA.65.033413}
  {\bibfield  {journal} {\bibinfo  {journal} {Phys. Rev. A}\ }\textbf {\bibinfo
  {volume} {65}},\ \bibinfo {pages} {033413} (\bibinfo {year}
  {2002})}\BibitemShut {NoStop}%
\bibitem [{\citenamefont {Vahala}\ \emph {et~al.}(2009)\citenamefont {Vahala},
  \citenamefont {Herrmann}, \citenamefont {Knünz}, \citenamefont {Batteiger},
  \citenamefont {Saathoff}, \citenamefont {Hänsch},\ and\ \citenamefont
  {Udem}}]{Vahala2009}%
  \BibitemOpen
  \bibfield  {author} {\bibinfo {author} {\bibfnamefont {K.}~\bibnamefont
  {Vahala}}, \bibinfo {author} {\bibfnamefont {M.}~\bibnamefont {Herrmann}},
  \bibinfo {author} {\bibfnamefont {S.}~\bibnamefont {Knünz}}, \bibinfo
  {author} {\bibfnamefont {V.}~\bibnamefont {Batteiger}}, \bibinfo {author}
  {\bibfnamefont {G.}~\bibnamefont {Saathoff}}, \bibinfo {author}
  {\bibfnamefont {T.~W.}\ \bibnamefont {Hänsch}},\ and\ \bibinfo {author}
  {\bibfnamefont {T.}~\bibnamefont {Udem}},\ }\href
  {https://doi.org/10.1038/nphys1367} {\bibfield  {journal} {\bibinfo
  {journal} {Nat. Phys.}\ }\textbf {\bibinfo {volume} {5}},\ \bibinfo {pages}
  {682} (\bibinfo {year} {2009})}\BibitemShut {NoStop}%
\bibitem [{\citenamefont {Schneider}\ \emph {et~al.}(2014)\citenamefont
  {Schneider}, \citenamefont {Schowalter}, \citenamefont {Chen}, \citenamefont
  {Sullivan},\ and\ \citenamefont {Hudson}}]{Schneider2014}%
  \BibitemOpen
  \bibfield  {author} {\bibinfo {author} {\bibfnamefont {C.}~\bibnamefont
  {Schneider}}, \bibinfo {author} {\bibfnamefont {S.~J.}\ \bibnamefont
  {Schowalter}}, \bibinfo {author} {\bibfnamefont {K.}~\bibnamefont {Chen}},
  \bibinfo {author} {\bibfnamefont {S.~T.}\ \bibnamefont {Sullivan}},\ and\
  \bibinfo {author} {\bibfnamefont {E.~R.}\ \bibnamefont {Hudson}},\ }\href
  {https://link.aps.org/doi/10.1103/PhysRevApplied.2.034013} {\bibfield
  {journal} {\bibinfo  {journal} {Phys. Rev. Applied}\ }\textbf {\bibinfo
  {volume} {2}},\ \bibinfo {pages} {034013} (\bibinfo {year}
  {2014})}\BibitemShut {NoStop}%
\bibitem [{\citenamefont {Deb}\ \emph {et~al.}(2015)\citenamefont {Deb},
  \citenamefont {Pollum}, \citenamefont {Smith}, \citenamefont {Keller},
  \citenamefont {Rennick}, \citenamefont {Heazlewood},\ and\ \citenamefont
  {Softley}}]{Deb2015}%
  \BibitemOpen
  \bibfield  {author} {\bibinfo {author} {\bibfnamefont {N.}~\bibnamefont
  {Deb}}, \bibinfo {author} {\bibfnamefont {L.~L.}\ \bibnamefont {Pollum}},
  \bibinfo {author} {\bibfnamefont {A.~D.}\ \bibnamefont {Smith}}, \bibinfo
  {author} {\bibfnamefont {M.}~\bibnamefont {Keller}}, \bibinfo {author}
  {\bibfnamefont {C.~J.}\ \bibnamefont {Rennick}}, \bibinfo {author}
  {\bibfnamefont {B.~R.}\ \bibnamefont {Heazlewood}},\ and\ \bibinfo {author}
  {\bibfnamefont {T.~P.}\ \bibnamefont {Softley}},\ }\href
  {https://doi.org/10.1103/PhysRevA.91.033408} {\bibfield  {journal} {\bibinfo
  {journal} {Phys. Rev. A}\ }\textbf {\bibinfo {volume} {91}},\ \bibinfo
  {pages} {033408} (\bibinfo {year} {2015})}\BibitemShut {NoStop}%
\bibitem [{\citenamefont {Schmid}\ \emph {et~al.}(2017)\citenamefont {Schmid},
  \citenamefont {Greenberg}, \citenamefont {Miller}, \citenamefont {Loeffler},\
  and\ \citenamefont {Lewandowski}}]{Schmid2017}%
  \BibitemOpen
  \bibfield  {author} {\bibinfo {author} {\bibfnamefont {P.~C.}\ \bibnamefont
  {Schmid}}, \bibinfo {author} {\bibfnamefont {J.}~\bibnamefont {Greenberg}},
  \bibinfo {author} {\bibfnamefont {M.~I.}\ \bibnamefont {Miller}}, \bibinfo
  {author} {\bibfnamefont {K.}~\bibnamefont {Loeffler}},\ and\ \bibinfo
  {author} {\bibfnamefont {H.~J.}\ \bibnamefont {Lewandowski}},\ }\href
  {https://doi.org/10.1063/1.4996911} {\bibfield  {journal} {\bibinfo
  {journal} {Rev. Sci. Instrum.}\ }\textbf {\bibinfo {volume} {88}},\ \bibinfo
  {pages} {123107} (\bibinfo {year} {2017})}\BibitemShut {NoStop}%
\bibitem [{\citenamefont {Staanum}\ \emph {et~al.}(2008)\citenamefont
  {Staanum}, \citenamefont {H\o{}jbjerre}, \citenamefont {Wester},\ and\
  \citenamefont {Drewsen}}]{Staanum2008}%
  \BibitemOpen
  \bibfield  {author} {\bibinfo {author} {\bibfnamefont {P.~F.}\ \bibnamefont
  {Staanum}}, \bibinfo {author} {\bibfnamefont {K.}~\bibnamefont
  {H\o{}jbjerre}}, \bibinfo {author} {\bibfnamefont {R.}~\bibnamefont
  {Wester}},\ and\ \bibinfo {author} {\bibfnamefont {M.}~\bibnamefont
  {Drewsen}},\ }\href {https://doi.org/10.1103/PhysRevLett.100.243003}
  {\bibfield  {journal} {\bibinfo  {journal} {Phys. Rev. Lett.}\ }\textbf
  {\bibinfo {volume} {100}},\ \bibinfo {pages} {243003} (\bibinfo {year}
  {2008})}\BibitemShut {NoStop}%
\bibitem [{\citenamefont {Willitsch}\ \emph {et~al.}(2008)\citenamefont
  {Willitsch}, \citenamefont {Bell}, \citenamefont {Gingell}, \citenamefont
  {Procter},\ and\ \citenamefont {Softley}}]{Willitsch2008}%
  \BibitemOpen
  \bibfield  {author} {\bibinfo {author} {\bibfnamefont {S.}~\bibnamefont
  {Willitsch}}, \bibinfo {author} {\bibfnamefont {M.~T.}\ \bibnamefont {Bell}},
  \bibinfo {author} {\bibfnamefont {A.~D.}\ \bibnamefont {Gingell}}, \bibinfo
  {author} {\bibfnamefont {S.~R.}\ \bibnamefont {Procter}},\ and\ \bibinfo
  {author} {\bibfnamefont {T.~P.}\ \bibnamefont {Softley}},\ }\href
  {https://doi.org/10.1103/PhysRevLett.100.043203} {\bibfield  {journal}
  {\bibinfo  {journal} {Phys. Rev. Lett.}\ }\textbf {\bibinfo {volume} {100}},\
  \bibinfo {pages} {043203} (\bibinfo {year} {2008})}\BibitemShut {NoStop}%
\bibitem [{\citenamefont {Gingell}\ \emph {et~al.}(2010)\citenamefont
  {Gingell}, \citenamefont {Bell}, \citenamefont {Oldham}, \citenamefont
  {Softley},\ and\ \citenamefont {Harvey}}]{Gingell2010}%
  \BibitemOpen
  \bibfield  {author} {\bibinfo {author} {\bibfnamefont {A.~D.}\ \bibnamefont
  {Gingell}}, \bibinfo {author} {\bibfnamefont {M.~T.}\ \bibnamefont {Bell}},
  \bibinfo {author} {\bibfnamefont {J.~M.}\ \bibnamefont {Oldham}}, \bibinfo
  {author} {\bibfnamefont {T.~P.}\ \bibnamefont {Softley}},\ and\ \bibinfo
  {author} {\bibfnamefont {J.~N.}\ \bibnamefont {Harvey}},\ }\href
  {https://doi.org/10.1063/1.3505142} {\bibfield  {journal} {\bibinfo
  {journal} {J. Chem. Phys.}\ }\textbf {\bibinfo {volume} {133}},\ \bibinfo
  {pages} {194302} (\bibinfo {year} {2010})},\ \Eprint
  {https://arxiv.org/abs/https://doi.org/10.1063/1.3505142}
  {https://doi.org/10.1063/1.3505142} \BibitemShut {NoStop}%
\bibitem [{\citenamefont {Goeders}\ \emph {et~al.}(2013)\citenamefont
  {Goeders}, \citenamefont {Clark}, \citenamefont {Vittorini}, \citenamefont
  {Wright}, \citenamefont {Viteri},\ and\ \citenamefont
  {Brown}}]{Goeders2013a}%
  \BibitemOpen
  \bibfield  {author} {\bibinfo {author} {\bibfnamefont {J.~E.}\ \bibnamefont
  {Goeders}}, \bibinfo {author} {\bibfnamefont {C.~R.}\ \bibnamefont {Clark}},
  \bibinfo {author} {\bibfnamefont {G.}~\bibnamefont {Vittorini}}, \bibinfo
  {author} {\bibfnamefont {K.}~\bibnamefont {Wright}}, \bibinfo {author}
  {\bibfnamefont {C.~R.}\ \bibnamefont {Viteri}},\ and\ \bibinfo {author}
  {\bibfnamefont {K.~R.}\ \bibnamefont {Brown}},\ }\href
  {https://doi.org/10.1021/jp312368a} {\bibfield  {journal} {\bibinfo
  {journal} {J. Phys. Chem. A}\ }\textbf {\bibinfo {volume} {117}},\ \bibinfo
  {pages} {9725} (\bibinfo {year} {2013})}\BibitemShut {NoStop}%
\bibitem [{\citenamefont {Groot-Berning}\ \emph {et~al.}(2019)\citenamefont
  {Groot-Berning}, \citenamefont {Stopp}, \citenamefont {Jacob}, \citenamefont
  {Budker}, \citenamefont {Haas}, \citenamefont {Renisch}, \citenamefont
  {Runke}, \citenamefont {Thörle-Pospiech}, \citenamefont {Düllmann},\ and\
  \citenamefont {Schmidt-Kaler}}]{Groot-Berning2019}%
  \BibitemOpen
  \bibfield  {author} {\bibinfo {author} {\bibfnamefont {K.}~\bibnamefont
  {Groot-Berning}}, \bibinfo {author} {\bibfnamefont {F.}~\bibnamefont
  {Stopp}}, \bibinfo {author} {\bibfnamefont {G.}~\bibnamefont {Jacob}},
  \bibinfo {author} {\bibfnamefont {D.}~\bibnamefont {Budker}}, \bibinfo
  {author} {\bibfnamefont {R.}~\bibnamefont {Haas}}, \bibinfo {author}
  {\bibfnamefont {D.}~\bibnamefont {Renisch}}, \bibinfo {author} {\bibfnamefont
  {J.}~\bibnamefont {Runke}}, \bibinfo {author} {\bibfnamefont
  {P.}~\bibnamefont {Thörle-Pospiech}}, \bibinfo {author} {\bibfnamefont
  {C.~E.}\ \bibnamefont {Düllmann}},\ and\ \bibinfo {author} {\bibfnamefont
  {F.}~\bibnamefont {Schmidt-Kaler}},\ }\href
  {https://link.aps.org/doi/10.1103/PhysRevA.99.023420} {\bibfield  {journal}
  {\bibinfo  {journal} {Phys. Rev. A}\ }\textbf {\bibinfo {volume} {99}},\
  \bibinfo {pages} {023420} (\bibinfo {year} {2019})}\BibitemShut {NoStop}%
\bibitem [{\citenamefont {Drewsen}\ \emph {et~al.}(2004)\citenamefont
  {Drewsen}, \citenamefont {Mortensen}, \citenamefont {Martinussen},
  \citenamefont {Staanum},\ and\ \citenamefont {S\o{}rensen}}]{Drewsen2004}%
  \BibitemOpen
  \bibfield  {author} {\bibinfo {author} {\bibfnamefont {M.}~\bibnamefont
  {Drewsen}}, \bibinfo {author} {\bibfnamefont {A.}~\bibnamefont {Mortensen}},
  \bibinfo {author} {\bibfnamefont {R.}~\bibnamefont {Martinussen}}, \bibinfo
  {author} {\bibfnamefont {P.}~\bibnamefont {Staanum}},\ and\ \bibinfo {author}
  {\bibfnamefont {J.~L.}\ \bibnamefont {S\o{}rensen}},\ }\href
  {https://doi.org/10.1103/PhysRevLett.93.243201} {\bibfield  {journal}
  {\bibinfo  {journal} {Phys. Rev. Lett.}\ }\textbf {\bibinfo {volume} {93}},\
  \bibinfo {pages} {243201} (\bibinfo {year} {2004})}\BibitemShut {NoStop}%
\bibitem [{\citenamefont {Janik}\ \emph {et~al.}(1985)\citenamefont {Janik},
  \citenamefont {Nagourney},\ and\ \citenamefont {Dehmelt}}]{Janik1985}%
  \BibitemOpen
  \bibfield  {author} {\bibinfo {author} {\bibfnamefont {G.}~\bibnamefont
  {Janik}}, \bibinfo {author} {\bibfnamefont {W.}~\bibnamefont {Nagourney}},\
  and\ \bibinfo {author} {\bibfnamefont {H.}~\bibnamefont {Dehmelt}},\ }\href
  {https://doi.org/10.1364/JOSAB.2.001251} {\bibfield  {journal} {\bibinfo
  {journal} {J. Opt. Soc. Am. B}\ }\textbf {\bibinfo {volume} {2}},\ \bibinfo
  {pages} {1251} (\bibinfo {year} {1985})}\BibitemShut {NoStop}%
\bibitem [{Sup()}]{Supplemental}%
  \BibitemOpen
  \href@noop {} {}\bibinfo {note} {See Supplemental Material, which includes
  Refs. \cite{Lett1989, Kemiktarak2014, Berkeland1998, Morigi2001,
  Kielpinski2000, Akerman2010}, for further discussion of ion motional
  amplification, secular frequency calculations, and OMS systematic
  effects}\BibitemShut {NoStop}%
\bibitem [{\citenamefont {Bl\"umel}\ \emph {et~al.}(1989)\citenamefont
  {Bl\"umel}, \citenamefont {Kappler}, \citenamefont {Quint},\ and\
  \citenamefont {Walther}}]{Bluemel1989}%
  \BibitemOpen
  \bibfield  {author} {\bibinfo {author} {\bibfnamefont {R.}~\bibnamefont
  {Bl\"umel}}, \bibinfo {author} {\bibfnamefont {C.}~\bibnamefont {Kappler}},
  \bibinfo {author} {\bibfnamefont {W.}~\bibnamefont {Quint}},\ and\ \bibinfo
  {author} {\bibfnamefont {H.}~\bibnamefont {Walther}},\ }\href
  {https://doi.org/10.1103/PhysRevA.40.808} {\bibfield  {journal} {\bibinfo
  {journal} {Phys. Rev. A}\ }\textbf {\bibinfo {volume} {40}},\ \bibinfo
  {pages} {808} (\bibinfo {year} {1989})}\BibitemShut {NoStop}%
\bibitem [{\citenamefont {Oberst}(1999)}]{Oberst1999}%
  \BibitemOpen
  \bibfield  {author} {\bibinfo {author} {\bibfnamefont {H.}~\bibnamefont
  {Oberst}},\ }\emph {\bibinfo {title} {Resonance fluorescence of single Barium
  ions}},\ \href
  {https://quantumoptics.at/images/publications/diploma/diplom_oberst.pdf}
  {Master's thesis},\ \bibinfo  {school} {University of Innsbruck} (\bibinfo
  {year} {1999})\BibitemShut {NoStop}%
\bibitem [{\citenamefont {Roßnagel}\ \emph {et~al.}(2015)\citenamefont
  {Roßnagel}, \citenamefont {Tolazzi}, \citenamefont {Schmidt-Kaler},\ and\
  \citenamefont {Singer}}]{Rossnagel2015}%
  \BibitemOpen
  \bibfield  {author} {\bibinfo {author} {\bibfnamefont {J.}~\bibnamefont
  {Roßnagel}}, \bibinfo {author} {\bibfnamefont {K.~N.}\ \bibnamefont
  {Tolazzi}}, \bibinfo {author} {\bibfnamefont {F.}~\bibnamefont
  {Schmidt-Kaler}},\ and\ \bibinfo {author} {\bibfnamefont {K.}~\bibnamefont
  {Singer}},\ }\href {http://dx.doi.org/10.1088/1367-2630/17/4/045004}
  {\bibfield  {journal} {\bibinfo  {journal} {New J. Phys.}\ }\textbf {\bibinfo
  {volume} {17}},\ \bibinfo {pages} {045004} (\bibinfo {year}
  {2015})}\BibitemShut {NoStop}%
\bibitem [{\citenamefont {Fan}\ \emph {et~al.}(2019)\citenamefont {Fan},
  \citenamefont {Holliman}, \citenamefont {Wang},\ and\ \citenamefont
  {Jayich}}]{Fan2019}%
  \BibitemOpen
  \bibfield  {author} {\bibinfo {author} {\bibfnamefont {M.}~\bibnamefont
  {Fan}}, \bibinfo {author} {\bibfnamefont {C.~A.}\ \bibnamefont {Holliman}},
  \bibinfo {author} {\bibfnamefont {A.~L.}\ \bibnamefont {Wang}},\ and\
  \bibinfo {author} {\bibfnamefont {A.~M.}\ \bibnamefont {Jayich}},\ }\href
  {https://link.aps.org/doi/10.1103/PhysRevLett.122.223001} {\bibfield
  {journal} {\bibinfo  {journal} {Phys. Rev. Lett.}\ }\textbf {\bibinfo
  {volume} {122}},\ \bibinfo {pages} {223001} (\bibinfo {year}
  {2019})}\BibitemShut {NoStop}%
\bibitem [{\citenamefont {Pruttivarasin}\ and\ \citenamefont
  {Katori}(2015)}]{Pruttivarasin2015}%
  \BibitemOpen
  \bibfield  {author} {\bibinfo {author} {\bibfnamefont {T.}~\bibnamefont
  {Pruttivarasin}}\ and\ \bibinfo {author} {\bibfnamefont {H.}~\bibnamefont
  {Katori}},\ }\href {https://doi.org/10.1063/1.4935476} {\bibfield  {journal}
  {\bibinfo  {journal} {Rev. Sci. Instrum.}\ }\textbf {\bibinfo {volume}
  {86}},\ \bibinfo {pages} {115106} (\bibinfo {year} {2015})}\BibitemShut
  {NoStop}%
\bibitem [{\citenamefont {Lu}\ and\ \citenamefont {Yang}(1998)}]{Lu1998}%
  \BibitemOpen
  \bibfield  {author} {\bibinfo {author} {\bibfnamefont {W.}~\bibnamefont
  {Lu}}\ and\ \bibinfo {author} {\bibfnamefont {S.}~\bibnamefont {Yang}},\
  }\href {https://doi.org/10.1021/jp9728969} {\bibfield  {journal} {\bibinfo
  {journal} {J. Phys. Chem. A}\ }\textbf {\bibinfo {volume} {102}},\ \bibinfo
  {pages} {825} (\bibinfo {year} {1998})}\BibitemShut {NoStop}%
\bibitem [{\citenamefont {Lee}\ and\ \citenamefont {Farrar}(2002)}]{Lee2002a}%
  \BibitemOpen
  \bibfield  {author} {\bibinfo {author} {\bibfnamefont {J.~I.}\ \bibnamefont
  {Lee}}\ and\ \bibinfo {author} {\bibfnamefont {J.~M.}\ \bibnamefont
  {Farrar}},\ }\href {https://doi.org/10.1021/jp0216979} {\bibfield  {journal}
  {\bibinfo  {journal} {J. Phys. Chem. A}\ }\textbf {\bibinfo {volume} {106}},\
  \bibinfo {pages} {11882} (\bibinfo {year} {2002})}\BibitemShut {NoStop}%
\bibitem [{\citenamefont {Puri}\ \emph {et~al.}(2017)\citenamefont {Puri},
  \citenamefont {Mills}, \citenamefont {Schneider}, \citenamefont {Simbotin},
  \citenamefont {Montgomery}, \citenamefont {Côté}, \citenamefont {Suits},\
  and\ \citenamefont {Hudson}}]{Puri2017}%
  \BibitemOpen
  \bibfield  {author} {\bibinfo {author} {\bibfnamefont {P.}~\bibnamefont
  {Puri}}, \bibinfo {author} {\bibfnamefont {M.}~\bibnamefont {Mills}},
  \bibinfo {author} {\bibfnamefont {C.}~\bibnamefont {Schneider}}, \bibinfo
  {author} {\bibfnamefont {I.}~\bibnamefont {Simbotin}}, \bibinfo {author}
  {\bibfnamefont {J.~A.}\ \bibnamefont {Montgomery}}, \bibinfo {author}
  {\bibfnamefont {R.}~\bibnamefont {Côté}}, \bibinfo {author} {\bibfnamefont
  {A.~G.}\ \bibnamefont {Suits}},\ and\ \bibinfo {author} {\bibfnamefont
  {E.~R.}\ \bibnamefont {Hudson}},\ }\href
  {http://science.sciencemag.org/content/early/2017/09/06/science.aan4701.abstract}
  {\bibfield  {journal} {\bibinfo  {journal} {Science}\ } (\bibinfo {year}
  {2017})}\BibitemShut {NoStop}%
\bibitem [{\citenamefont {Augenbraun}\ \emph {et~al.}(2020)\citenamefont
  {Augenbraun}, \citenamefont {Lasner}, \citenamefont {Frenett}, \citenamefont
  {Sawaoka}, \citenamefont {Miller}, \citenamefont {Steimle},\ and\
  \citenamefont {Doyle}}]{Augenbraun2020}%
  \BibitemOpen
  \bibfield  {author} {\bibinfo {author} {\bibfnamefont {B.~L.}\ \bibnamefont
  {Augenbraun}}, \bibinfo {author} {\bibfnamefont {Z.~D.}\ \bibnamefont
  {Lasner}}, \bibinfo {author} {\bibfnamefont {A.}~\bibnamefont {Frenett}},
  \bibinfo {author} {\bibfnamefont {H.}~\bibnamefont {Sawaoka}}, \bibinfo
  {author} {\bibfnamefont {C.}~\bibnamefont {Miller}}, \bibinfo {author}
  {\bibfnamefont {T.~C.}\ \bibnamefont {Steimle}},\ and\ \bibinfo {author}
  {\bibfnamefont {J.~M.}\ \bibnamefont {Doyle}},\ }\href
  {https://doi.org/10.1088/1367-2630/ab687b} {\bibfield  {journal} {\bibinfo
  {journal} {New J. Phys.}\ }\textbf {\bibinfo {volume} {22}},\ \bibinfo
  {pages} {022003} (\bibinfo {year} {2020})}\BibitemShut {NoStop}%
\bibitem [{Cou()}]{Coursey2015}%
  \BibitemOpen
  \href@noop {} {}\bibinfo {note} {Coursey, J.S., Schwab, D.J., Tsai, J.J., and
  Dragoset, R.A. (2015), Atomic Weights and Isotopic Compositions (version
  4.1). [Online] Available: http://physics.nist.gov/Comp [2020 Mar 11].
  National Institute of Standards and Technology, Gaithersburg,
  MD.}\BibitemShut {Stop}%
\bibitem [{\citenamefont {Khanyile}\ \emph {et~al.}(2015)\citenamefont
  {Khanyile}, \citenamefont {Shu},\ and\ \citenamefont {Brown}}]{Khanyile2015}%
  \BibitemOpen
  \bibfield  {author} {\bibinfo {author} {\bibfnamefont {N.~B.}\ \bibnamefont
  {Khanyile}}, \bibinfo {author} {\bibfnamefont {G.}~\bibnamefont {Shu}},\ and\
  \bibinfo {author} {\bibfnamefont {K.~R.}\ \bibnamefont {Brown}},\ }\href
  {https://doi.org/10.1038/ncomms8825} {\bibfield  {journal} {\bibinfo
  {journal} {Nat. Commun.}\ }\textbf {\bibinfo {volume} {6}},\ \bibinfo {pages}
  {7825} (\bibinfo {year} {2015})}\BibitemShut {NoStop}%
\bibitem [{\citenamefont {Schmidt}\ \emph {et~al.}(2005)\citenamefont
  {Schmidt}, \citenamefont {Rosenband}, \citenamefont {Langer}, \citenamefont
  {Itano}, \citenamefont {Bergquist},\ and\ \citenamefont
  {Wineland}}]{Schmidt2005}%
  \BibitemOpen
  \bibfield  {author} {\bibinfo {author} {\bibfnamefont {P.~O.}\ \bibnamefont
  {Schmidt}}, \bibinfo {author} {\bibfnamefont {T.}~\bibnamefont {Rosenband}},
  \bibinfo {author} {\bibfnamefont {C.}~\bibnamefont {Langer}}, \bibinfo
  {author} {\bibfnamefont {W.~M.}\ \bibnamefont {Itano}}, \bibinfo {author}
  {\bibfnamefont {J.~C.}\ \bibnamefont {Bergquist}},\ and\ \bibinfo {author}
  {\bibfnamefont {D.~J.}\ \bibnamefont {Wineland}},\ }\href
  {http://science.sciencemag.org/content/309/5735/749.abstract} {\bibfield
  {journal} {\bibinfo  {journal} {Science}\ }\textbf {\bibinfo {volume}
  {309}},\ \bibinfo {pages} {749} (\bibinfo {year} {2005})}\BibitemShut
  {NoStop}%
\bibitem [{\citenamefont {Wolf}\ \emph {et~al.}(2016)\citenamefont {Wolf},
  \citenamefont {Wan}, \citenamefont {Heip}, \citenamefont {Gebert},
  \citenamefont {Shi},\ and\ \citenamefont {Schmidt}}]{Wolf2016}%
  \BibitemOpen
  \bibfield  {author} {\bibinfo {author} {\bibfnamefont {F.}~\bibnamefont
  {Wolf}}, \bibinfo {author} {\bibfnamefont {Y.}~\bibnamefont {Wan}}, \bibinfo
  {author} {\bibfnamefont {J.~C.}\ \bibnamefont {Heip}}, \bibinfo {author}
  {\bibfnamefont {F.}~\bibnamefont {Gebert}}, \bibinfo {author} {\bibfnamefont
  {C.}~\bibnamefont {Shi}},\ and\ \bibinfo {author} {\bibfnamefont {P.~O.}\
  \bibnamefont {Schmidt}},\ }\href {http://dx.doi.org/10.1038/nature16513}
  {\bibfield  {journal} {\bibinfo  {journal} {Nature}\ }\textbf {\bibinfo
  {volume} {530}},\ \bibinfo {pages} {457} (\bibinfo {year}
  {2016})}\BibitemShut {NoStop}%
\bibitem [{\citenamefont {Pal}\ \emph {et~al.}(2009)\citenamefont {Pal},
  \citenamefont {Jiang}, \citenamefont {Safronova},\ and\ \citenamefont
  {Safronova}}]{Pal2009}%
  \BibitemOpen
  \bibfield  {author} {\bibinfo {author} {\bibfnamefont {R.}~\bibnamefont
  {Pal}}, \bibinfo {author} {\bibfnamefont {D.}~\bibnamefont {Jiang}}, \bibinfo
  {author} {\bibfnamefont {M.~S.}\ \bibnamefont {Safronova}},\ and\ \bibinfo
  {author} {\bibfnamefont {U.~I.}\ \bibnamefont {Safronova}},\ }\href
  {https://link.aps.org/doi/10.1103/PhysRevA.79.062505} {\bibfield  {journal}
  {\bibinfo  {journal} {Phys. Rev. A}\ }\textbf {\bibinfo {volume} {79}},\
  \bibinfo {pages} {062505} (\bibinfo {year} {2009})}\BibitemShut {NoStop}%
\bibitem [{\citenamefont {Ratschbacher}\ \emph {et~al.}(2012)\citenamefont
  {Ratschbacher}, \citenamefont {Zipkes}, \citenamefont {Sias},\ and\
  \citenamefont {Köhl}}]{Ratschbacher2012}%
  \BibitemOpen
  \bibfield  {author} {\bibinfo {author} {\bibfnamefont {L.}~\bibnamefont
  {Ratschbacher}}, \bibinfo {author} {\bibfnamefont {C.}~\bibnamefont
  {Zipkes}}, \bibinfo {author} {\bibfnamefont {C.}~\bibnamefont {Sias}},\ and\
  \bibinfo {author} {\bibfnamefont {M.}~\bibnamefont {Köhl}},\ }\href
  {https://doi.org/10.1038/nphys2373} {\bibfield  {journal} {\bibinfo
  {journal} {Nat. Phys.}\ }\textbf {\bibinfo {volume} {8}},\ \bibinfo {pages}
  {649} (\bibinfo {year} {2012})}\BibitemShut {NoStop}%
\bibitem [{\citenamefont {Johnson}\ \emph {et~al.}(2016)\citenamefont
  {Johnson}, \citenamefont {Wong-Campos}, \citenamefont {Restelli},
  \citenamefont {Landsman}, \citenamefont {Neyenhuis}, \citenamefont
  {Mizrahi},\ and\ \citenamefont {Monroe}}]{Johnson2016}%
  \BibitemOpen
  \bibfield  {author} {\bibinfo {author} {\bibfnamefont {K.~G.}\ \bibnamefont
  {Johnson}}, \bibinfo {author} {\bibfnamefont {J.~D.}\ \bibnamefont
  {Wong-Campos}}, \bibinfo {author} {\bibfnamefont {A.}~\bibnamefont
  {Restelli}}, \bibinfo {author} {\bibfnamefont {K.~A.}\ \bibnamefont
  {Landsman}}, \bibinfo {author} {\bibfnamefont {B.}~\bibnamefont {Neyenhuis}},
  \bibinfo {author} {\bibfnamefont {J.}~\bibnamefont {Mizrahi}},\ and\ \bibinfo
  {author} {\bibfnamefont {C.}~\bibnamefont {Monroe}},\ }\href
  {http://iontrap.umd.edu/wp-content/uploads/2016/05/rf_stabilization.pdf}
  {\bibfield  {journal} {\bibinfo  {journal} {Rev. Sci. Instrum.}\ } (\bibinfo
  {year} {2016})}\BibitemShut {NoStop}%
\bibitem [{\citenamefont {Lett}\ \emph {et~al.}(1989)\citenamefont {Lett},
  \citenamefont {Phillips}, \citenamefont {Rolston}, \citenamefont {Tanner},
  \citenamefont {Watts},\ and\ \citenamefont {Westbrook}}]{Lett1989}%
  \BibitemOpen
  \bibfield  {author} {\bibinfo {author} {\bibfnamefont {P.~D.}\ \bibnamefont
  {Lett}}, \bibinfo {author} {\bibfnamefont {W.~D.}\ \bibnamefont {Phillips}},
  \bibinfo {author} {\bibfnamefont {S.~L.}\ \bibnamefont {Rolston}}, \bibinfo
  {author} {\bibfnamefont {C.~E.}\ \bibnamefont {Tanner}}, \bibinfo {author}
  {\bibfnamefont {R.~N.}\ \bibnamefont {Watts}},\ and\ \bibinfo {author}
  {\bibfnamefont {C.~I.}\ \bibnamefont {Westbrook}},\ }\href
  {http://josab.osa.org/abstract.cfm?URI=josab-6-11-2084} {\bibfield  {journal}
  {\bibinfo  {journal} {J. Opt. Soc. Am. B}\ }\textbf {\bibinfo {volume} {6}},\
  \bibinfo {pages} {2084} (\bibinfo {year} {1989})}\BibitemShut {NoStop}%
\bibitem [{\citenamefont {Kemiktarak}\ \emph {et~al.}(2014)\citenamefont
  {Kemiktarak}, \citenamefont {Durand}, \citenamefont {Metcalfe},\ and\
  \citenamefont {Lawall}}]{Kemiktarak2014}%
  \BibitemOpen
  \bibfield  {author} {\bibinfo {author} {\bibfnamefont {U.}~\bibnamefont
  {Kemiktarak}}, \bibinfo {author} {\bibfnamefont {M.}~\bibnamefont {Durand}},
  \bibinfo {author} {\bibfnamefont {M.}~\bibnamefont {Metcalfe}},\ and\
  \bibinfo {author} {\bibfnamefont {J.}~\bibnamefont {Lawall}},\ }\href
  {https://doi.org/10.1103/PhysRevLett.113.030802} {\bibfield  {journal}
  {\bibinfo  {journal} {Phys. Rev. Lett.}\ }\textbf {\bibinfo {volume} {113}},\
  \bibinfo {pages} {030802} (\bibinfo {year} {2014})}\BibitemShut {NoStop}%
\bibitem [{\citenamefont {Berkeland}\ \emph {et~al.}(1998)\citenamefont
  {Berkeland}, \citenamefont {Miller}, \citenamefont {Bergquist}, \citenamefont
  {Itano},\ and\ \citenamefont {Wineland}}]{Berkeland1998}%
  \BibitemOpen
  \bibfield  {author} {\bibinfo {author} {\bibfnamefont {D.~J.}\ \bibnamefont
  {Berkeland}}, \bibinfo {author} {\bibfnamefont {J.~D.}\ \bibnamefont
  {Miller}}, \bibinfo {author} {\bibfnamefont {J.~C.}\ \bibnamefont
  {Bergquist}}, \bibinfo {author} {\bibfnamefont {W.~M.}\ \bibnamefont
  {Itano}},\ and\ \bibinfo {author} {\bibfnamefont {D.~J.}\ \bibnamefont
  {Wineland}},\ }\href {https://doi.org/http://dx.doi.org/10.1063/1.367318}
  {\bibfield  {journal} {\bibinfo  {journal} {J. Appl. Phys.}\ }\textbf
  {\bibinfo {volume} {83}},\ \bibinfo {pages} {5025} (\bibinfo {year}
  {1998})}\BibitemShut {NoStop}%
\bibitem [{\citenamefont {Morigi}\ and\ \citenamefont
  {Walther}(2001)}]{Morigi2001}%
  \BibitemOpen
  \bibfield  {author} {\bibinfo {author} {\bibfnamefont {G.}~\bibnamefont
  {Morigi}}\ and\ \bibinfo {author} {\bibfnamefont {H.}~\bibnamefont
  {Walther}},\ }\href {https://doi.org/10.1007/s100530170275} {\bibfield
  {journal} {\bibinfo  {journal} {Eur. Phys. J. D}\ }\textbf {\bibinfo {volume}
  {13}},\ \bibinfo {pages} {261} (\bibinfo {year} {2001})}\BibitemShut
  {NoStop}%
\bibitem [{\citenamefont {Kielpinski}\ \emph {et~al.}(2000)\citenamefont
  {Kielpinski}, \citenamefont {King}, \citenamefont {Myatt}, \citenamefont
  {Sackett}, \citenamefont {Turchette}, \citenamefont {Itano}, \citenamefont
  {Monroe}, \citenamefont {Wineland},\ and\ \citenamefont
  {Zurek}}]{Kielpinski2000}%
  \BibitemOpen
  \bibfield  {author} {\bibinfo {author} {\bibfnamefont {D.}~\bibnamefont
  {Kielpinski}}, \bibinfo {author} {\bibfnamefont {B.~E.}\ \bibnamefont
  {King}}, \bibinfo {author} {\bibfnamefont {C.~J.}\ \bibnamefont {Myatt}},
  \bibinfo {author} {\bibfnamefont {C.~A.}\ \bibnamefont {Sackett}}, \bibinfo
  {author} {\bibfnamefont {Q.~A.}\ \bibnamefont {Turchette}}, \bibinfo {author}
  {\bibfnamefont {W.~M.}\ \bibnamefont {Itano}}, \bibinfo {author}
  {\bibfnamefont {C.}~\bibnamefont {Monroe}}, \bibinfo {author} {\bibfnamefont
  {D.~J.}\ \bibnamefont {Wineland}},\ and\ \bibinfo {author} {\bibfnamefont
  {W.~H.}\ \bibnamefont {Zurek}},\ }\href
  {https://doi.org/10.1103/PhysRevA.61.032310} {\bibfield  {journal} {\bibinfo
  {journal} {Phys. Rev. A}\ }\textbf {\bibinfo {volume} {61}},\ \bibinfo
  {pages} {032310} (\bibinfo {year} {2000})}\BibitemShut {NoStop}%
\bibitem [{\citenamefont {Akerman}\ \emph {et~al.}(2010)\citenamefont
  {Akerman}, \citenamefont {Kotler}, \citenamefont {Glickman}, \citenamefont
  {Dallal}, \citenamefont {Keselman},\ and\ \citenamefont
  {Ozeri}}]{Akerman2010}%
  \BibitemOpen
  \bibfield  {author} {\bibinfo {author} {\bibfnamefont {N.}~\bibnamefont
  {Akerman}}, \bibinfo {author} {\bibfnamefont {S.}~\bibnamefont {Kotler}},
  \bibinfo {author} {\bibfnamefont {Y.}~\bibnamefont {Glickman}}, \bibinfo
  {author} {\bibfnamefont {Y.}~\bibnamefont {Dallal}}, \bibinfo {author}
  {\bibfnamefont {A.}~\bibnamefont {Keselman}},\ and\ \bibinfo {author}
  {\bibfnamefont {R.}~\bibnamefont {Ozeri}},\ }\href
  {https://doi.org/10.1103/PhysRevA.82.061402} {\bibfield  {journal} {\bibinfo
  {journal} {Phys. Rev. A}\ }\textbf {\bibinfo {volume} {82}},\ \bibinfo
  {pages} {061402(R)} (\bibinfo {year} {2010})}\BibitemShut {NoStop}%
\end{thebibliography}
\end{document}


\title{Supplemental material}

\maketitle

\renewcommand\thefigure{S\arabic{figure}} 
\renewcommand\theequation{S\arabic{equation}} 
\renewcommand\thetable{S\arabic{table}} 

\section{Ion motion amplification}

In an 1D harmonic oscillator with normal frequency $\omega_0$, the position and velocity of a single ion at time $t$ are

\begin{align}
    x(t) &= x_0 \cos{\omega_0 t}, \\
    v(t) &= x_0\omega_0 \sin{\omega_0 t},
\label{eq:motion}
\end{align}
where $x_0$ is the motion amplitude.

A light field with $k$-vector $\vec{k}_L$ couples to the ion motion through photon scatterings. We define $\vec{k}_L\cdot\hat{x}<0$, and the light force on the ion at time $t$ is

\begin{equation}
\label{eq:force-scatter}
    F(t) = F(\omega) = s(\omega)\hbar c \vec{k}_L\cdot\hat{x},
\end{equation}
where $s(\omega)$ is the scattering rate at Doppler shifted light frequency $\omega=\omega(v(t)) = c k_L\sqrt{(c+v(t))/(c-v(t))} \approx \omega_L(1+v(t)/c)$, where $\omega_L=c k_L$ is the light frequency in the lab frame.

From the work-energy theorem, the energy gained by the ion after a cycle of motion is

\begin{equation}
\label{eq:work-energy-x}
    \Delta E = \int_{x_0}^{-x_0}F'(x)\mathrm{d}x + \int_{-x_0}^{x_0}F''(x)\mathrm{d}x,
\end{equation}
where $F'(x)$ is the light force as a function of ion position in the first half of the cycle ($0<t<\frac{\pi}{\omega_0}$), and $F''(x)$ is the light force in the second half of the cycle ($\frac{\pi}{\omega_0}<t<\frac{2\pi}{\omega_0}$). By changing the integration variable in Eq. \ref{eq:work-energy-x} from $x$ to $t$,

\begin{equation}
\label{eq:work-energy-t}
    \Delta E = \int_{0}^{\frac{\pi}{\omega_0}} F(\omega)v(t)\mathrm{d}t + \int_{\frac{\pi}{\omega_0}}^{\frac{2\pi}{\omega_0}} F(\omega)v(t)\mathrm{d}t.
\end{equation}

From Eq. \ref{eq:motion},

\begin{equation}
\label{eq:v-time-reverse}
    v(t) = -v(\frac{2\pi}{\omega_0}-t).
\end{equation}

With this, Eq. \ref{eq:work-energy-t} becomes,

\begin{equation}
\label{eq:work-energy-noncombined}
\begin{split}
    \Delta E &= \int_{0}^{\frac{\pi}{\omega_0}} F(\omega(v(t)))v(t)\mathrm{d}t + \int_{\frac{\pi}{\omega_0}}^{\frac{2\pi}{\omega_0}} F(\omega(-v(\frac{2\pi}{\omega_0}-t)))(-v(\frac{2\pi}{\omega_0}-t))\mathrm{d}t\\
    &= \int_{0}^{\frac{\pi}{\omega_0}} F(\omega(v(t)))v(t)\mathrm{d}t + \int_{\frac{2\pi}{\omega_0}}^{\frac{\pi}{\omega_0}} F(\omega(-v(\frac{2\pi}{\omega_0}-t)))v(\frac{2\pi}{\omega_0}-t)\mathrm{d}t.
\end{split}
\end{equation}

We define $t'=\frac{2\pi}{\omega_0}-t$, and

\begin{equation}
\label{eq:work-energy}
\begin{split}
    \Delta E &= \int_{0}^{\frac{\pi}{\omega_0}} F(\omega(v(t)))v(t)\mathrm{d}t - \int_{0}^{\frac{\pi}{\omega_0}} F(\omega(-v(t')))v(t')\mathrm{d}t'\\
    &= \int_{0}^{\frac{\pi}{\omega_0}} [F(\omega(v(t))) - F(\omega(-v(t)))]v(t)\mathrm{d}t.
\end{split}
\end{equation}

From Eq. \ref{eq:force-scatter},

\begin{equation}
\label{eq:delta_e}
\Delta E = \hbar c \vec{k}_L\cdot\hat{x}\int_{0}^{\frac{\pi}{\omega_0}} [s(\omega(v)) - s(\omega(-v))]v(t)\mathrm{d}t.
\end{equation}

If $\Delta E > 0$, the ion's motion will be amplified after a cycle. If $\Delta E < 0$, the motional amplitude decreases after a cycle until reaching the Doppler cooling limit \cite{Lett1989}. From Eq. \ref{eq:delta_e}, the difference between scattering rates $s(\omega(v))$ and $s(\omega(-v))$ for $0 < t \leq \frac{\pi}{\omega_0}$  determines the sign of $\Delta E$.

Note that for small oscillation amplitude $x_0$ so that the scattering rate $s(\omega)$ is linear with $\omega$,

\begin{equation}
\label{eq:delta_e_slope}
\Delta E \propto \lim_{v\rightarrow0}[s(\omega(v)) - s(\omega(-v))] \propto -\left.\frac{\mathrm{d}s}{\mathrm{d}v}\right|_{0} \propto -\left.\frac{\mathrm{d}s}{\mathrm{d}\omega}\right|_{\omega_L},
\end{equation}
which gives the well-known result that the ion is cooled (heated) when the sign of the spectrum slope is positive (negative).

For spectra with multiple peaks, as in Fig. 2 in the main text, the ion is locally heated as the sign of the spectrum slope at $\omega = \omega_L$ is positive, but is globally cooled due to the contribution of light that is Doppler shifted when the ion motion is large that $\omega - \omega_L$ is no longer small compared to sizes of the CPT features. Therefore, the ion's motion is amplified up to a stable orbit due to an equilibrium between ``local heating'' and ``global cooling'' effects.

\begin{figure}[h]
    \centering
    \includegraphics{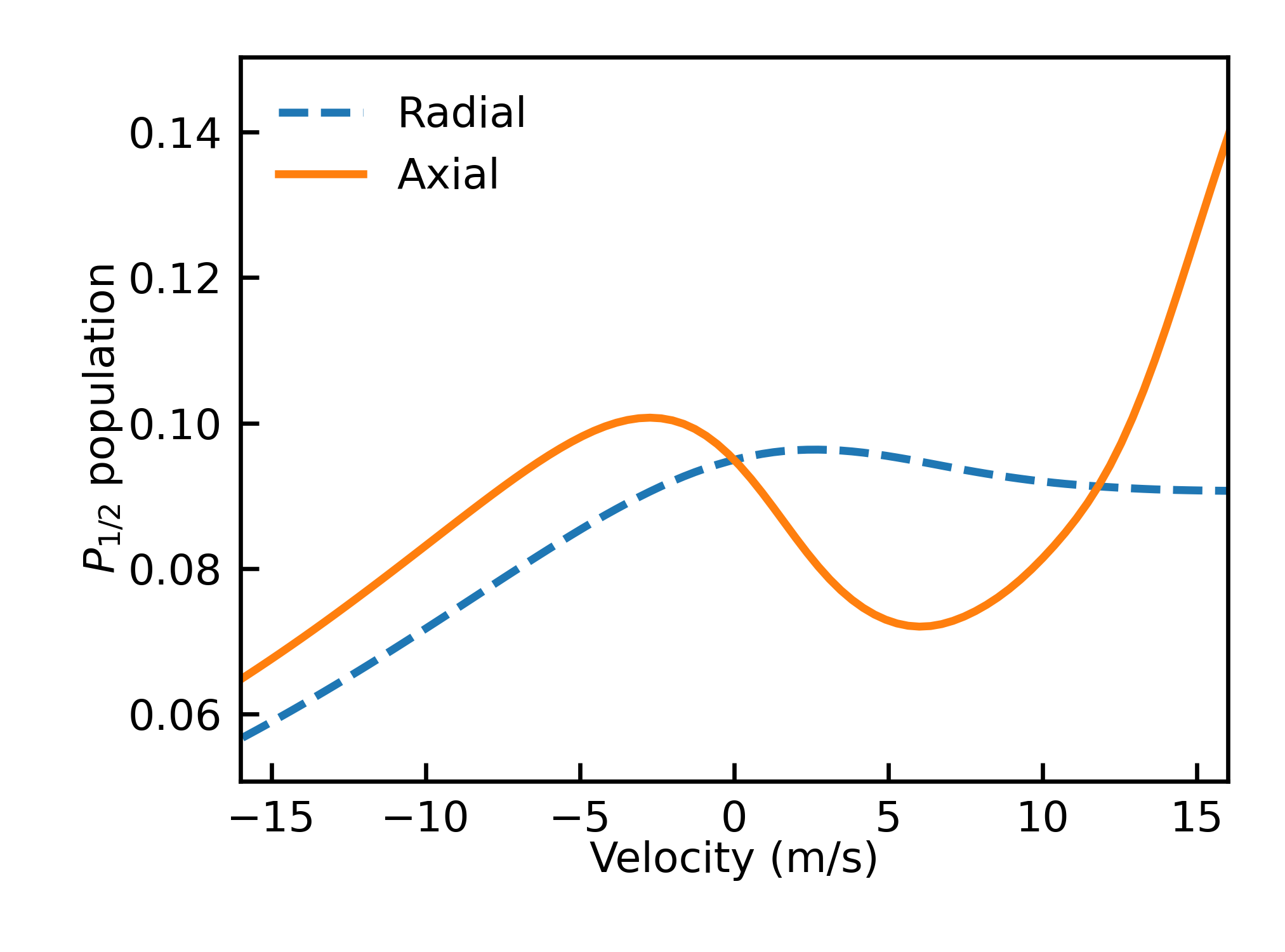}
    \caption{$P_{1/2}$ state population vs. radial and axial ion velocities. From the ion velocity we can calculate the Doppler shifted 468 nm and 1079 nm light frequencies, and the $P_{1/2}$ state population can be calculated from the light frequencies. The positive axes are defined so $\vec{k}_{468}\cdot\hat{x}<0$.}
    \label{fig:vel}
\end{figure}

We further analyze the ion's motion in the trap radial and axial directions given the experimental setup in Fig. 1 described in the main text. The 468 nm and 1079 nm $k$-vectors are perpendicular to each other, leading to anisotropy in axial and radial motion if we consider Doppler shifts to both the 468 nm and 1079 nm light: When the ion moves radially, the 468 and 1079 nm Doppler shifts have the same sign, and when the ion moves axially, the Doppler shifts have the opposite signs. We calculate the $P_{1/2}$ state population separately for ion velocities in the radial and axial directions, see Fig. \ref{fig:vel}, using the fitted parameters in Fig. 2 of the main text. According to Eq. \ref{eq:delta_e_slope}, the ion is cooled if the slope of the scattering rate, which is proportional to the $P_{1/2}$ state population, at $v=0$ is positive (radial directions), and the ion is heated if the slope of the scattering rate at $v=0$ is negative (axial direction). Therefore, the ion motion is selectively amplified in the axial direction.

We note that if multiple modes can be excited, mode competition will lead to stable amplification of only one mode \cite{Kemiktarak2014}, and the amplified coherent motion of the dominant mode can be used in OMS to measure ion masses.

\section{Secular frequencies of a linear ion crystal}

Formulas for the axial center-of-mass (COM) mode of a linear ion crystal with 1 to 3 ions are summarized.

We define the axial Mathieu parameter \cite{Berkeland1998}

\begin{equation}
    a_z = \frac{8Q\kappa U_0}{m {z_0}^2 {\Omega_{\mathrm{rf}}}^2},
\end{equation}
where $Q$ is the ion's charge, $m$ is the ion mass, $\kappa$ is a dimensionless factor related to shielding of the axial electric field by the radial electrodes, $U_0$ is the dc voltage on the two endcap electrodes, and $z_0$ is the distance from the endcap electrode to the trap center.

For a single ion, the axial secular frequency in the approximation of $a_z\ll 1$ is \cite{Berkeland1998}

\begin{equation}
    \omega_{z, 1} = \frac{\Omega_{\mathrm{rf}}}{2}\sqrt{a_z}.
\end{equation}

For a linear 2-ion crystal with ion masses $M$ and $m$, the axial COM mode frequency is \cite{Morigi2001}

\begin{equation}
    \omega_{z, 2} = \omega_{z, 1} \sqrt{1+\frac{1}{\mu}-\sqrt{1+\frac{1}{\mu^2}-\frac{1}{\mu}}},
\label{eq:z2}
\end{equation}
where $\mu$ is the mass ratio $M/m$.

For a linear 3-ion crystal with two ions of mass $m$ on each end, and an ion of mass $M$ at the center, the axial COM mode frequency is \cite{Kielpinski2000}

\begin{equation}
    \omega_{z, 3} = \omega_{z, 1} \sqrt{\frac{13}{10}+\frac{1}{10\mu}(21-\sqrt{441-34\mu+169\mu^2})}.
\label{eq:z3}
\end{equation}

\section{OMS systematics}

\subsection{Mass calibration}

We calculate the molecular ion masses using an OMS calibration measurement directly before introducing reactants, and the systematic shift is the difference between the ion mass using the most recent calibration data and that using the initial calibration shown in Fig. 3 in the main text. We do not report an uncertainty for this systematic as all calibrations have the same statistical uncertainty.

\subsection{Trap potential drift}

\begin{figure}[h]
    \centering
    \includegraphics{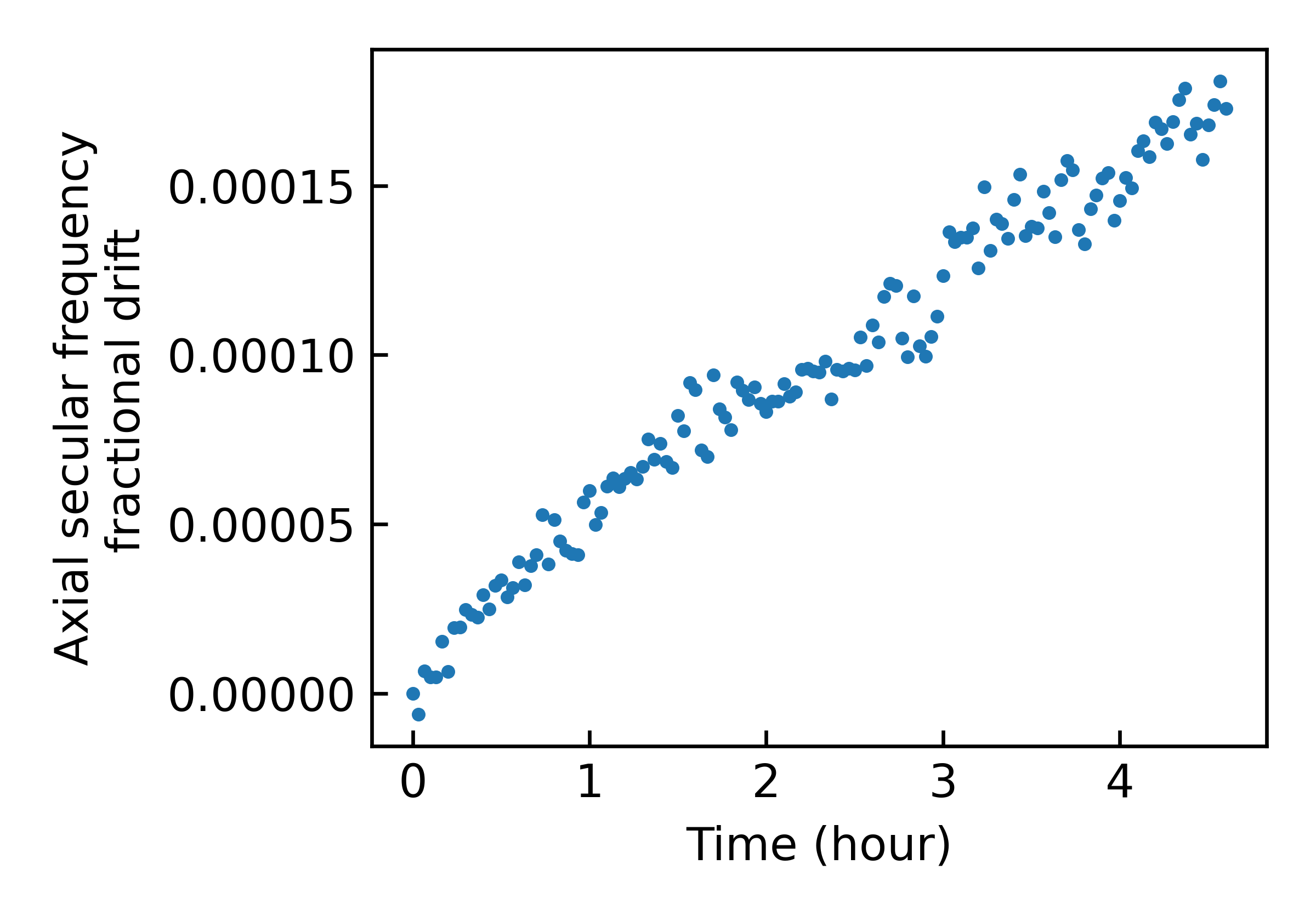}
    \caption{Fractional drift in the axial secular frequency of a single  $\mathrm{Ra}^+$.}
    \label{fig:drift}
\end{figure}

The secular frequency may drift in the time between mass calibration and molecular mass spectrometry (typically $<$ 1 hour), due to for example material deposited on the trap electrodes during ion loading. We measure the axial secular frequency of a single $\mathrm{Ra}^+$ as a function of time using the OMS for a period of $\sim5$ hours to measure the drift of the trapping potential, and the results are shown in Fig. \ref{fig:drift}. During the measurement, the maximum fractional frequency drift during an 1 hour period is $\SI{6e-5}{}$, which we use for the systematic error due to trap potential drift. The corresponding mass uncertainties are calculated.

\subsection{Secular motion amplitude shift}

\begin{figure}[h]
    \centering
    \includegraphics{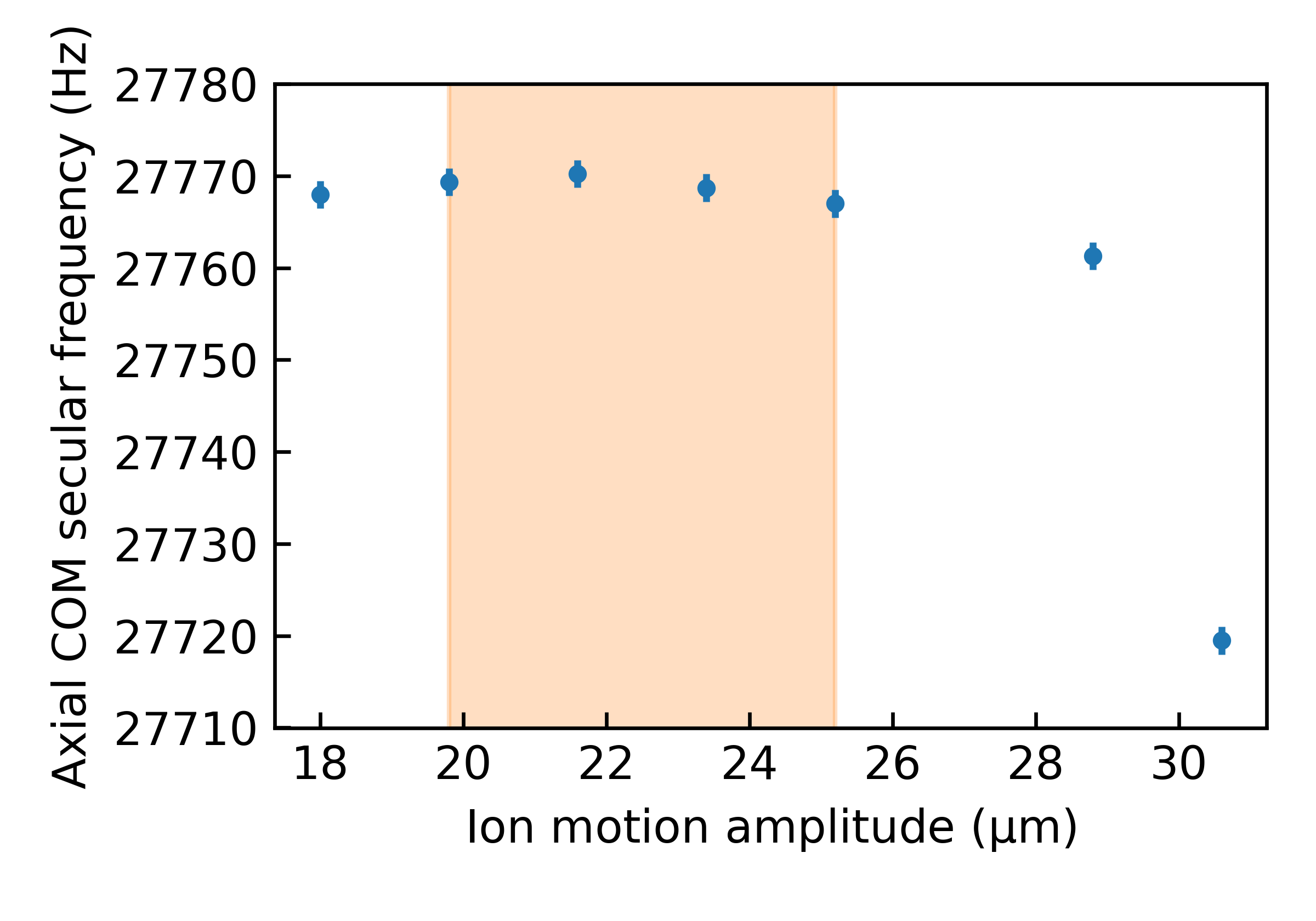}
    \caption{Secular frequency vs. ion secular motion amplitude for 3 $\mathrm{Ra}^+$. The orange shaded region indicates secular motion amplitudes used for OMS measurements.}
    \label{fig:amp}
\end{figure}

The axial COM secular frequency is amplitude-dependent due to trap anharmonicity \cite{Akerman2010}. We measure the axial COM mode secular frequency for a 3 $\mathrm{Ra}^+$ crystal as a function of the oscillation amplitude [See Fig. \ref{fig:amp}]. With a motional amplitude of $\SI{22\pm3}{\micro\meter}$ that is used in the measurement, the maximum fractional secular frequency shift due to oscillation amplitude uncertainty is $\SI{5e-5}{}$, which translates to mass uncertainties that are less than 0.1 dalton.

The systematic shifts and uncertainties of measured masses are summarized in Table \ref{tab:mass}, along with the statistical results. The total systematic shift is a linear sum of the individual shifts, and the total uncertainty is a quadrature sum of the individual uncertainties.

\begin{table}[]
    \centering
    \begin{ruledtabular}
    \begin{tabular}{lcccc}
         & \ce{RaOH+} & \ce{RaOD+} & \ce{RaOCH3+} & \ce{RaOCD3+} \\
        \hline \\ [-0.6pc]
        Calibration shift & 0.01 & 0.21 & 0.34 & 0.22 \\
        Trap potential drift & 0.00(9) & 0.00(9) & 0.00(9) & 0.00(9) \\
        Motion amplitude shift & 0.00(7) & 0.00(7) & 0.00(8) & 0.00(8) \\
        \hline \\ [-1.2pc]
        Total systematic & 0.01(11) & 0.21(11) & 0.34(11) & 0.22(11)
    \end{tabular}
    \end{ruledtabular}
    \caption{Systematic shifts and uncertainties of the measured molecular ion masses in daltons.}
    \label{tab:mass}
\end{table}